\documentclass{article}
\usepackage{graphicx}
\usepackage{amssymb,latexsym}

\begin{document}

\title{Action  Quantization, Energy Quantization, \\
and Time Parametrization}

\author{Edward R Floyd \\
10 Jamaica Village Road, Coronado, CA 92118-3208, USA \\
Email: floyd@mailaps.org}

\date{\today}

\maketitle

\begin{abstract}
The additional information within a Hamilton-Jacobi representation of quantum mechanics is extra, in general, to the Schr\"odinger representation. This additional information specifies the microstate of $\psi$ that is incorporated into the quantum reduced action, $W$. Non-physical solutions of the quantum stationary Hamilton-Jacobi equation  for energies that are not Hamiltonian eigenvalues are examined to establish Lipschitz continuity of the quantum reduced action and conjugate momentum.  Milne quantization renders the eigenvalue $J$.  Eigenvalues $J$ and $E$ mutually imply each other.   Jacobi's theorem generates a microstate-dependent time parametrization $t-\tau=\partial_E W$ even where energy, $E$, and action variable, $J$, are quantized eigenvalues.  Substantiating examples are examined in a Hamilton-Jacobi representation including the linear harmonic oscillator numerically and the square well in closed form. Two byproducts are developed.  First, the monotonic behavior of $W$ is shown to ease numerical and analytic computations. Second, a Hamilton-Jacobi representation, quantum trajectories, is shown to develop the standard energy quantization formulas of wave mechanics.
\end{abstract}

\small


\medskip

\noindent PhySH:  quantum foundations,  1 dimensional systems,  Ab initio calculation

\normalsize

\section{Introduction}

The additional information available in a Hamilton-Jacobi formulation of quantum mechanics [\ref{bib:prd25}--\ref{bib:ijmpa15}], and extra to the $\psi$ representation, can be used to develop insight into the foundations of quantum mechanics.  Quantum trajectories representation (QT) of quantum mechanics  [\ref{bib:prd25}--\ref{bib:rc},\ref{bib:wyatt}] couched in its underlying Hamilton-Jacobi formulation, is used herein for time parametrization in an open-space universe.  The open-universe assumption was tacit in previous QT calculations.  An investigation of a spinless anyon in one dimension, $q$, is sufficient to achieve time parametrization.  The quantum reduced action (Hamilton's quantum characteristic function), $W$, is the generater of motion for the quantum trajectory.  The quantum reduced action is described by the quantum stationary Hamilton-Jacobi equation (QSHJE).  The QSHJE is a phenomenological third-order partial differential equation expressed in one Cartesian dimension $q$ by

\begin{equation}
\frac{(\partial W/\partial q)^2}{2m} + V(q) - E = -\frac{\hbar^2}{4m}\underbrace{\langle W;q \rangle}_{\mbox{\scriptsize Schwarzian derivative}}
\label{eq:qshje}
\end{equation}

\noindent where $V$ is the potential, $m$ is mass, $\hbar$ is Planck's constant, and $\langle W;q \rangle$ is the Schwarzian derivative of $W$
with respect to $q$.  If the left side of Eq.\ (\ref{eq:qshje}) by itself were equal to zero, then it would represent the classical Hamilton-Jacobi equation (CSHJE).  The Schwarzian derivative contains higher-order derivatives that manifest the quantum effects and is given by

\begin{equation}
\langle W;q \rangle = \frac{\partial^3 W/\partial q^3}{\partial W/\partial q} - \frac{3}{2} \left( \frac{\partial^2 W/\partial q^2}{\partial W/\partial q}
\right)^2.
\label{eq:sd}
\end{equation}

\noindent The conjugate momentum, $\partial_q W$ is also a solution of the QSHJE, for $W$ does not explicitly appear in the QSHJE,

The quantum stationary Hamilton-Jacobi equation (QSHJE) is a non-linear, third-order partial differential equation while the stationary Schr\"odinger equation (SSE) is a linear second-order partial differential equation.  The classical stationary Hamilton-Jacobi equation (CSHJE) is a non-linear first-order partial differential equation.   Hence, the solution to the QSHJE, the quantum reduced action (Hamilton's quantum characteristic function), $W$ contains more information than the Schr\"odinger wave function, $\psi$,  [\ref{bib:prd25}--\ref{bib:fm}] as well as the classical reduced action, $W_{\mbox{\scriptsize classical}}$.  This additional information is the higher order derivatives (additional initial values at $q_0$ for the third-order QSHJE where $q_0$ is the initial point) that specify a unique solution, $W(\{W,\partial_q W,{\partial^2}_q W\}|_{q=q_0};E,q)$, of the QSHJE.  For a bound state, the energy eigenvalue by itself is insufficient to specify the quantum reduced action; the initial conditions are also needed to do so.  Yet, any particular reduced action of energy eigenvalue $E$ (a microstate) is sufficient to specify the solution $\psi$ of the associated Schr\"odinger equation [\ref{bib:prd25}--\ref{bib:fm}].  Similarly, a unique solution for the conjugate momentum of the QSHJE is specified by $\partial_q W(\{\partial_q W,{\partial^2}_q W\}|_{q=q_0};E,q)$.

Under stationarity, the relationship between the quantum Hamilton's principal function $S$ and $W$ by QT may be given by

\begin{equation}
S(\{S,\partial_q S,{\partial^2}_q S\}|_{q=q_0};t,q) = W(\{\partial_q W,{\partial^2}_q W\}|_{q=q_0};E,q) - E(t-\tau)
\label{eq:sw}
\end{equation}

\noindent  in units of action.  Under stationarity, the initial values $\{S,\partial_q S,{\partial^2}_q S\}|_{q=q_0} = \{W,\partial_q W,{\partial^2}_q W\}|_{q=q_0}$, and $\partial_t S = -E$ where $\tau$ is the constant coordinate specifying the epoch (the time at the finite initial point $q_0$).  As $Et$ is the product in units of action of a conjugate pair of variables, Eq.\ (\ref{eq:sw}) is reminiscent of the switching of independent variable of a conjugate pair of variables by a Legendre transform (here the switch for QT is from variable $t$ in $S$ to variable $E$ in $W$).  When QT manipulates $W$, a generator of quantum motion, such as in Jacobi's theorem, $E$ is a continuous variable of $W$ during the manipulation.  The results of the manipulation can then be evaluated for some specified particular value of $E$ that becomes a constant of the motion.  For a particular bound state, the specification would be $E_{\mbox{\scriptsize particular eigenvalue}}$.

In the QT representation, quantization of energy, $E$, for bound states was developed by Milne quantization where the quantum action variable is quantized [\ref{bib:pr35}],

\begin{equation}
J_n \equiv J(E_n) = \oint \partial_q W(\{\partial_q W,{\partial^2}_q W\}|_{q=q_0};E_n,q) \, dq = n2\pi\hbar,\ \ \ \ n=1,2,3,\cdots,
\label{eq:jquant}
\end{equation}

\noindent without consideration whether the corresponding Schr\"odinger wave function, $\psi$, is $L^2(\mathbb{R})$ [\ref{bib:prd25},\ref{bib:prvda25}] nor whether $\psi$ represents a Bornian probability amplitude [\ref{bib:prd25}--\ref{bib:prd34}].  Nevertheless $\psi$ being $L^2(\mathbb{R})$ is a consequence of $J$ quantization. The energy $E_n$ is the energy for the $n$th bound state.  Note that $J_n$ is independent of initial values $\{\partial_q W,{\partial^2}_q W\}|_{q=q_0}$ for $\partial_q W$.  Even though $J_n$ and $E_n$ mutually imply each other, knowledge of both together still cannot remove the afore-mentioned arbitrariness of the initial conditions for $\partial_q W$ or $W$.

The closed integration path for $J_n$ in Eq.\ (\ref{eq:jquant}) may be executed either numerically or analytically in closed form.  The closed form integration is general and has been presented in detail in Ref.\ \ref{bib:prd34} by a contour integration in the complex plane of $\phi(q)/\vartheta(q)$ where a convenient change of variables has been executed by $q \mapsto \phi(q)/\vartheta(q)$ and where $\{\phi(q),\vartheta(q)\}$ is the set of independent solutions of the associated Schr\"odinger equation with $\vartheta(q)$ being the bound-state solution. This mapping changes the integrand to algebraic. The contour integration shows in closed form that $J_n$ is independent of the particular microstate of the conjugate momentum.  Numerical integration formally calculates $J_n = W|^{+\infty}_{q=-\infty} - W|^{-\infty}_{q=+\infty}$ numerically where $W|^{+\infty}_{q=-\infty}$ is on one side of a cut with terminals at $q = \pm \infty$ and $W|^{-\infty}_{q=+\infty}$ is on the other side.

Faraggi and Matone have developed the remarkable quantum equivalence principle (QEP) for which one-particle systems reversibly map into each other by coordinate transformations [\ref{bib:fm},\ref{bib:pl450}].   QEP implies the same quantized spectrum of the Copenhagen interpretation.   The quantum equivalence principle is reconciled with the Copenhagen interpretation by deriving the Copenhagen interpretation's main axioms by a simple geometric formulation [\ref{bib:fm}]. QT and QEP derive the same quantum reduced action, $W$, for the same initial values from the same QSHJE.  For bound states, QEP like QT finds $W$ not to be uniquely specified by an energy eigenvalues but needs the initial values $\{W,\partial_q W,{\partial^2}_q W\}|_{q=q_0}$, which again yields microstates (M\"obius states as used by Faraggi and Matone) [\ref{bib:prd26}--\ref{bib:fm}].  This additional information is extra to the $\psi$ representation of quantum mechanics [\ref{bib:prd29}--\ref{bib:fm}].

In classical mechanics, Jacobi's theorem renders time parametrization. Also QT parametrizes time by Jacobi's theorem in an open universe.  Jacobi's theorem is a critical step in the QT algorithm for time parametrization.  Jacobi's theorem for quantum mechanics for stationarity may be expressed as

\begin{eqnarray}
t - \tau & = & \frac{\partial W(\{W,\partial_q W,{\partial^2}_q W\}|_{q=q_0};E,q)}{\partial E} \nonumber \\
         & \equiv & \left. \left( \frac{\partial W(\{W,\partial_q W,{\partial^2}_q W\}|_{E=E_{\mbox{\tiny eigenvalue}},q=q_0};E_{\mbox{\scriptsize variable}},q)}{\partial E_{\mbox{\scriptsize variable}}} \right) \right|_{E_{\mbox{\scriptsize variable}} = E_{\mbox{\scriptsize eigenvalue}}}
\label{eq:jacobitheorem}
\end{eqnarray}

\noindent  where the right side of the first line is the convenient representation of the application Jacobi's theorem applied to quantum reduced action and the second line describes the application more precisely, albeit more cumbersomely.  In the first line of Eq.\ (\ref{eq:jacobitheorem}),  $t$ is time, $\tau$ is the constant coordinate that sets the epoch, $W$ is the quantum reduced action (a generator of nonlocal quantum motion), $E$ is the energy, and $\{W,\partial_q W,{\partial^2}_q W\}|_{q=q_0}$ is the set of initial values at $q_0$ necessary and sufficient to solve the QSHJE uniquely for the reduced action with energy $E$.  The set of initial values also specifies the particular microstate for the quantum reduced action.   In the second line, $E_{\mbox{\scriptsize variable}}$ explicitly denotes that energy is a variable per Eq.\ (\ref{eq:sw}) and  $\{W,\partial_q W,{\partial^2}_q W\}|_{E=E_{\mbox{\tiny eigenvalue}},q=q_0}$ means the set of initial values have values specified by $E_{\mbox{\scriptsize eigenvalue}},q=q_0$ so that the initial values are constants and not subject to change under differentiation by $E_{\mbox{\scriptsize variable}}$.  That is when considering the continuity of quantum reduced action with small changes of energy for the differentiation process, the initial values remain fixed constants. The initial values for $W$ are compatible with nil variational restrictions placed upon the trajectory end points of the associated Lagrangian as discussed further in \S2.2.

Jacobi's theorem generates an equation of deterministic, nonlocal quantum motion. As $W$ has the dimension of action and the relationship between quantum Hamilton's principal function $S(t,q)$ and $W(E,q)$ is given for stationarity by Eq.\ (\ref{eq:sw}), Jacobi's theorem is reminiscent of the switching of independent variable of a conjugate pair of variables of a Legendre transform (here the switch has been done from variable $t$ in $S$ to variable $E$ in $W$).  Equation (\ref{eq:jacobitheorem}) may be derived by partial differentiating Eq.\ (\ref{eq:sw}) by $E$.  The equation of quantum motion describes a deterministic, nonlocal quantum trajectory in space-time that is microstate dependent. (Herein ``deterministic" implies that the particular initial values, $\{W,\partial_q W,{\partial^2}_q W\}|_{q=q_0}$ for $W$  render a unique quantum trajectory for a given $E$ and $q_0$.  Different initial values would render other unique quantum trajectories for the same $E$ and $q_0$.  Different quantum trajectories for the same $E$ and $q_0$ imply different microstates.)  The additional information, that is incorporated into the quantum reduced action, is unavailable in the $\psi$ representation.

Faraggi and Matone have reservations about the validity of applying Jacobi's theorem to the quantum reduced action to generate local quantum trajectories and to parametrize time [\ref{bib:fm3}--\ref{bib:fm5}].  For completeness, practitioners of the Bohmian interpretation of quantum mechanics have expressed unpublished similar reservations about using Jacobi's theorem due to the quantization of energy.  While QEP and QT solve the QSHJE for the same quantum action variable, Faraggi and Matone reason that if space were compact, then bound states as well as free states would only have discrete (quantized) energy spectra that would preclude time parametrization by Jacobi's theorem.  Compactness of the universe is presently an open question of cosmology (whose answer is beyond the scope of this opus).  While Faraggi and Matone do consider Jacobi's theorem to be a good semi-classical approximation, they still consider fundamentally that the very existence of quantum trajectories would contradict the Bornian probabilistic axiom of the Copenhagen interpretation of quantum mechanics where ``the concept of localized particle with a defined velocity does not exist" [\ref{bib:fm3}].  Again, QT uses the additional information inherent in $W$ to generate a family of unique quantum trajectories that cover space.  Each microstate by itself implies $\psi$ [\ref{bib:prd25}--\ref{bib:prd34}].

Both the QSHJE and SSE are solvable for non-Hamilton eigenvalue energies.  These {\itshape ab initio} solutions are not eigenfunctions.  Still, non-eigenfunction solutions are useful in this opus to establish Lipschitz continuity with respect to energy of Hamilton-Jacobi algorithm in general and the quantum reduced action in particular.  Again, the time variable in Hamilton's quantum principal function is switched to the energy variable in the quantum reduced action as implied by Eq.\ (\ref{eq:sw}).  In this opus, if value of energy used in QSHJE or Schr\"odinger equation is not the Hamiltonian eigenvalue, then such energy is denoted by $\widetilde{E}$.  $\widetilde{W}$ and $\widetilde{\psi}$ are respectively non-eigenfunction solutions to the QSHJE and Schr\"odinger equation for non-eigenvalue energy $\widetilde{E}$. Non-eigenvalues and non-eigenfunctions are described herein by the adjective ``virtual".  Here, ``virtual" denotes not physically existing per se, but an {\itshape ab initio} computation may make it appear to be so without any attempt to deceive.   By precept, $\widetilde{E}$ is piecewise continuous in an open energy domain between limit points of eigenvalue $E$.  Equation (\ref{eq:jacobitheorem}) may be generalized for virtual energies by substituting to $\widetilde{E}$ for $ E_{\mbox{\scriptsize eigenvalue}}$.

The goal of this opus is to show that QT can parametrize time for an open universe.  QT will realize this goal by accomplishing the following tasks in reverse order.  Time parametrization, in general, is established if it can be done for bound states.  QT utilizes Jacobi's theorem for time parametrization.  Jacobi's theorem differentiates the quantum reduced action by energy. This differentiation will be shown to be``well posed" in the Hadamard sense even for discreet energy eigenvalues.  To be a well posed computation, its solution must (1) exist, (2) be unique, and (3) be Lipschitz continuous (i.e., small changes in $E$ or initial values will result in small changes in time or the algorithm output) [\ref{bib:jh},\ref{bib:ik}].  As quantum reduced action is differentiated by energy in Jacobi's theorem by $E$, it must be Lipschitz continuous in energy over its neighborhood of discreet energy of interest.  The phenomenological QSHJE for the quantum reduced action will be shown to be well posed not only for the discreet $E$ of interest but also for neighboring virtual $\widetilde{E}$s as well. QT will then show that $\lim_{\widetilde{E} \to E_{\mbox{\tiny eigenvalue}}} \widetilde{W}(\widetilde{E},q) = W(E_{\mbox{\scriptsize eigenvalue}},q)$.  Then Jacobi's theorem for bound states will be well posed for

\begin{equation}
\lim_{\widetilde{E}\to E_{\mbox{\tiny eigenvalue}}} \ \left( \frac{\partial \widetilde{W}(\{W,\partial_q W,{\partial^2}_q W\}|_{q=q_0};\widetilde{E},q)}{\partial \widetilde{E}}\right) = \frac{\partial W(\{W,\partial_q W,{\partial^2}_q W\}|_{q=q_0};E,q)}{\partial E_{\mbox{\scriptsize eigenvalue}}},
\label{eq:limjacobitheorem}
\end{equation}

\noindent  where the initial values $\{W,\partial_q W,{\partial^2}_q W\}|_{q=q_0}$ are those for eigenfunction $W$ for $E_{\mbox{\scriptsize eigenvalue}}$.  Time parametrization that is microstate-dependent will follow where in turn $\lim_{\widetilde{E}\to E_{\mbox{\tiny eigenvalue}}} (t-\widetilde{\tau}) = t-\tau_{\mbox{\scriptsize eigenvalue}}$ by  Eqs.\ (\ref{eq:jacobitheorem}) and (\ref{eq:limjacobitheorem}).  Application will be made to linear harmonic oscillator numerically and to the finite square well in closed form.

Section 2 applies Jacobi's theorem numerically to the linear harmonic oscillator.  Time parametrization is shown not to be unique just for given energy but also dependent on the initial values for the quantum reduced action. Even virtual states are shown to obey the Bohr correspondence principle. In \S 3, Jacobi's theorem is applied to the finite square well.  Closed-form solutions as functions of familiar elementary transcendental functions are developed for quantum reduced actions and time parametrizations. Quantizing formulas are developed.  Generalizations to other potentials are briefly examined.  In \S 4, findings and conclusions are discussed.

\section{Linear Harmonic Oscillator}

The quantization of the quantum action variable for the linear harmonic oscillator was first investigated numerically in 1982 [\ref{bib:prd25},\ref{bib:prvda25}].  The monotonic behavior of the quantum reduced action [\ref{bib:fm}] makes it a better candidate for numerical analyses than $\psi$.  Herein, the QSHJE for the linear harmonic oscillator is again examined numerically by the Mathcad\textsuperscript{\texttrademark} routine Rkadapt to greater accuracy (15 significant figures) and in more depth than it was  in 1982.  Rkadapt executes an adaptively spaced Runge-Kutta algorithm of fifth order that helps to suppress parasitic solutions deep in the classically forbidden region.
Rkadapt adjusts spacing with the behavior of the solution but returns equally-spaced values of the solution.  Rkadapt is used to solve the QSHJE for the quantum reduced
action, $W$, and the conjugate momentum, $\partial W/\partial q$, as $W$ does not explicitly appear in the QSHJE, Eqs.\ (\ref{eq:qshje}) and (\ref{eq:sd}).

\subsection{Energy and Action Variable}

The set of initial values for the third-order QSHJE for a given energy are $\{W,\partial_q W,{\partial^2}_q W\}|_{q=q_0}$ where $q_0$ is an initial point located for computational convenience at the mirror-symmetric point (a stable point) of the potential well for the linear harmonic oscillator  [\ref{bib:prd29}].   For completeness, the set of initial values may be given by either $\{W,\partial_qW,{\partial^2}_qW\}|_{q=q_0}$ or, among other possibilities, by $\{q,\dot{q},\ddot{q}\}|_{q=q_0}$, [\ref{bib:prd29}].  Faraggi and Matone have developed the relationships between $\{W,\partial_q W,{\partial^2}_q W\}$ and $\{q,\dot{q},\ddot{q}\}$ [\ref{bib:fm}].

The quantum reduced action is a generator of quantum motion.  Each particular quantum reduced action is a particular microstate of the bound state $\psi$ of eigenvalue energy $E$ [\ref{bib:fpl9},\ref{bib:fm}].  Each $W(\{W,\partial_q W,{\partial^2}_q W/\}|_{q=q_0};E,q)$ generates its particular
trajectory on the $q,t$-plane by Jacobi's theorem.  Thus, many quantum trajectories describe microstates of the same bound-state $\psi$ [\ref{bib:fpl9},\ref{bib:fm}]. A virtual quantum trajectory may still be generated from the virtual quantum reduced action for a virtual (non-eigenvalue) energy $\widetilde{E}$ by Jacobi's theorem.  Time parametrization for a quantum trajectory is innate to generating the quantum trajectory in the $q,t$-plane.  (Hamilton-Jacobi theory in one dimension is set in the $q,t$-plane.)  We generate time parametrization for the linear harmonic oscillator herein by a numerical calculation of Jacobi's theorem by finite differences.

The potential for the linear harmonic oscillator is given by $V(q)=m\omega^2q^2/2$ where $\omega^2$ is the force constant and $\omega$ is the classical frequency of the linear harmonic oscillator.  Natural units  where $\omega=1,\ m=1,$ and $\hbar=1$ are used herein \S2 where computations are done numerically.  We solve the QSHJE, Eq.\ (\ref{eq:qshje}), for the linear harmonic oscillator for
$E=0.4,0.5,0.6,\cdots,1.4,1.5,1.6$.  This energy range includes the ground state energy, $E=0.5$, and the energy for the first excited state, $E=1.5$,
in natural units.  The other energies are virtual (non-eigenvalue), $\widetilde{E}$, for which the phenomenological QSHJE renders $\widetilde{W}$ while the SSE renders virtual states, $\widetilde{\psi}$.  For computing simplicity, we choose quantum reduced actions that are antisymmetric about $q=0$.  This is achieved by setting the initial point to be $q_0 = 0$ with the initial value for $({\partial^2}_q W)|_{q=0} = 0$.  This achieves a symmetric conjugate momentum and an antisymmetric quantum reduced action if $W(0)=0$.  Rkadapt uses 4000 points herein to solve the QSHJE for $W$ in the range $0 \le q \le 10$ (natural units).  For each selected value of energy, $E$, we have chosen various initial values for conjugate momentum in natural
units of momentum given by $(\partial_q W)|_{q=0} = 0.5,(2mE)^{1/2},1,2$ to calculate $W(E,q)$ in natural units of action.  The initial conjugate
momenta are $0.5$ for Case A, $1$ for Case C, and $2$ for Case D.  Cases A, C, and D specify the same initial values of that Case for the various energies.  The initial value of the conjugate momentum $(2mE)^{1/2}$ for Case B does not maintain the same microstate during a variation in energy, for its initial values change with energy by precept (to be discussed further in \S2.2 and \S3).  Case B is used to study the rate of confirmation of Bohr's correspondence tendencies as the discrete eigenvalue energy increases from level ``$n$" to ``$n+1$" and also even for virtual energies from ``$n-1/2$" to ``$n+1/2$".

Case C is the most intuitive case among Cases A, C, and D for investigating the ground state of the linear harmonic oscillator.  The ground-state quantum reduced action for Case C best mimics the classical reduced action of the classical linear harmonic oscillator for which $V = m\omega^2 q^2/2$ with $E=0.5$ in natural units.  Let us compare, for the ground state, quantum and classical motions.  The classical and quantum reduced actions may have their integration constants selected so that both reduced actions are nil at the initial point $q_0=0$, where the potential energy is nil, $V = m\omega^2 q^2/2|_{q=0}=0$, for both the QSHJE and CSHJE.  The classical conjugate momentum is given by

\[
\frac{\partial W_{\mbox{\scriptsize classical}}}{\partial q} = (2mE-m\omega^2q^2)^{1/2} = (2E-q^2)^{1/2}\ \ \mbox{in natural units of momentum}.
\]

\noindent At $q=0$ the classical conjugate momentum analogy to the quantum ground state ($E_{\mbox{\scriptsize gs}}=1/2$) is given by $(2E)^{1/2} = 1$ in natural units.  The positive sign is assumed consistent with the direction of propagation.  Case C also has its initial value for conjugate momentum given by $\partial_q W|_{q=0}=1$, so that the initial classical and quantum conjugate momenta are equal.  As $q=0$ is also the equilibrium point for the classical linear harmonic oscillator, its acceleration, $\ddot{q}$, at $q=0$ must be nil. The derivative of the classical conjugate momentum at $q=0$ is given by

\[
\frac{\partial^2 W_{\mbox{\scriptsize classical}}}{\partial q^2}\bigg|_{q=0} = \frac{-m\omega^2q}{(2mE-m\omega^2 q^2)^{-1/2}}\bigg|_{q=0} = 0.
\]

\noindent For Case C, ${\partial^2}_q W|_{q=0}=0$ is another specified initial value for the quantum reduced action. The second derivative of the classical conjugate momentum at $q=0$ is given by

\[
\frac{\partial^3 W_{\mbox{\scriptsize classical}}}{\partial q^3}\bigg|_{q=0} = \frac{-m\omega^2}{(2mE)^{1/2}} = -(2E)^{-1/2}\ \ \mbox{in natural units}.
\]

\noindent  For $E=0.5$, we have ${\partial^3}_q W_{\mbox{\scriptsize classical}} = -1$.  On the other hand for the quantum motion, ${\partial^3}_q W|_{q=0}$ is extra
to the set of initial values $\{W,\partial_q W,{\partial^2}_q W\}|_{q=q_0}$ for solving the QSHJE uniquely.  For quantum motion, ${\partial^3}_q W|_{q=0}$ is dependent upon the set of initial values, $\{W,\partial_q W,{\partial^2}_q W\}|_{q=q_0}$ and the QSHJE itself.  For Case C where $\partial_q W|_{q=0} = 1$, all energy, $E$, for the ground state is allocated at the initial point ($q=0$) to the kinetic energy term, $(\partial_q W)^2/(2m)$.  Again, Case C has nil potential energy at the initial point $V=m\omega^2q^2/2|_{q=0}=0$.  By the QSHJE, nil energy can thus be allocated to balance the quantum term, $\hbar^2 \langle W;q \rangle/(4m)$, at the initial point. Within the quantum term at $q=0$, the two terms of the Schwarzian derivative, Eq.\ (\ref{eq:sd}), must be in balance with each other with regard to energy.  The second term on the right side of Eq.\ (\ref{eq:sd}) is nil for $({\partial^2}_q W)|_{q=0} = 0$ by  the initial values of Case C.  It follows that the first term of the Schwarzian derivative must also be nil at the initial point leaving $({\partial^3}_q W)|_{q=0} = 0$ for Case C.  For Case C and CSHJE,  $({\partial^3}_q W)|_{q=0}$ and $({\partial^3}_q W_{\mbox{\scriptsize classical}})|_{q=0}$ differ.  As a result for Case C, the classical and quantum reduced actions form a crossing osculation at $q=0$ for ground-state energy.  Nevertheless, for $E=(2n-1)/2,\ n=1,2,3,\cdots$ in natural units
$J_{\mbox{\scriptsize classical}}=(2n-1)\pi$ while in the quantum environment $J=2n\pi$.  On the other hand, the quantum reduced actions for Cases A and D
cross in a non-tangential manner with the classical reduced action at $q=0$.

Numerically calculated values of the action variable, $J$, or virtual action variables, $\widetilde{J}$, for various Cases and various energies are presented in
Table 1. The (virtual) action variables are approximated by symmetry by $J \approx 4W(10)$ and $\widetilde{J} \approx 4\widetilde{W}(10)$ where the initial value of
(virtual) quantum action variable is set at $W(0)=0$ and $\widetilde{W}(0)=0$. The symmetry allows us to examine only a quarter-cycle of the action variable.
The validity of this approximation is substantiated by Figures 1 and 2 where not only does the conjugate momenta attenuate rapidly in the classical forbidden region
but also the product of the conjugate momenta and the divergence factor, the reciprocal Gaussian factor $\exp(+q^2/2)$.  The attenuation with $q$ of the product of the conjugate momenta and the divergence factor also substantiates rigorously the Lipschitz continuity in $q$ of the conjugate momentum, $\partial_qW$, which is also a solution to the QSHJE.  Figure 1 examines the attenuation of the conjugate momentum for the ground state for Case C; Figure 2, for the non-eigenvalue $\widetilde{E}=1$.  The numerical attenuation of the (virtual) conjugate momentum, $\partial_q W$ or $\partial_q \widetilde{W}$ is reported to become smaller than computational round-off of Rkadapt at the range $6\le q < 7$ (natural units).  Such attenuation reduces the (virtual) conjugate momentum to many orders smaller than could be discerned visually on Figures 1 and 2.  The rapid attenuation of the (virtual) conjugate momentum down to the Rkadapt computer limit before $q = 7$ justifies the approximations $J \approx 4W(10)$ and $\widetilde{J} \approx 4\widetilde{W}(10)$ where $4W(5)$ and $4\widetilde{W}(5)$ are represented by the areas under the solid curves on Figs.\ 1 and 2.

\begin{table}[t]
\begin{center}
\caption{\small Numerically calculated (virtual) quantum action variables, $J$ for eigenvalues and $\widetilde{J}$ for non-eigenvalues, in natural units for the linear harmonic oscillator. Calculations are done over a range of (virtual) energy, $E$ and $\widetilde{E}$, spanning beyond the ground state and first excited state and are done for various cases specified by the initial value of conjugate momentum at the origin $(q=0)$.  For all cases, the other initial values at the origin are $W=0$ and $\partial^2 W/\partial q^2 = 0$. The tabulated values of $J\omega/(2\pi) - E - 0.5\hbar \omega$ or $J/(2\pi) - E -0.5$ in natural units for and only for eigenstates are almost nil and being of the order $10^{-15}$ in natural units. As such, the values of $J/(2\pi)-E-0.5$ in natural units approach the computational accuracy of Rkadapt for eigenvalue $J$s.  For non-eigenvalues of $\widetilde{E}$, the associated $\widetilde{J}$ exhibits finite values that are dependent upon energy and the initial values. Presentation of values are in natural units.}

\begin{tabular}{lr|llll}
\hline
                      &                               &                      &                   &                    &                   \\[-6pt]
                      & Case:                         &   \ A                &  \ \,B            &    \ C             & \ D              \\
                 &$(\partial W/\partial q)|_{q=0}$\ = & $0.5$                & $(2E)^{1/2}$      & $\ 1$              & $\ 2$    \\[3pt] \hline
                      &                               &                      &                   &                    &                   \\[-6pt]
  $\widetilde{E}=0.4$ & $\widetilde{J} \approx$       & $\ 1.592 \pi$        & $1.766 \pi$       & $1.791 \pi$        & $1.895 \pi$     \\
  \ \ \ "             & $\widetilde{J}/(2\pi)-E-0.5=$ & $\ -0.104$           & $-0.0168$         & $-0.00473$         & $+0.0473$          \\[6pt]
  $E=0.5$             & $J \approx$                   & $\ 2 \pi$            & $2 \pi$           & $2 \pi$            & $2 \pi$         \\
  \ \ \ "             & $|J/(2\pi)-E-0.5|<$           & $\ 2\times10^{-15}$  &$2\times10^{-15}$  & $2\times10^{-15}$  &$2\times10^{-15}$   \\[6pt]
  $\widetilde{E}=0.6$ & $\widetilde{J} \approx$       & $\ 2.464 \pi$        & $2.220 \pi$       & $2.240 \pi$        & $2.121 \pi$     \\
  \ \ \ "             & $\widetilde{J}/(2\pi)-E-0.5=$ & $\ +0.132$           & $+0.00979$        & $+0.0200$          & $-0.0395$         \\[6pt]
  $\widetilde{E}=0.7$ & $\widetilde{J} \approx$       & $\ 2.882 \pi6$       & $2.430 \pi$       & $2.501 \pi$        & $2.261 \pi$    \\
  \ \ \ "             & $\widetilde{J}/(2\pi)-E-0.5=$ & $\ +0.241$           & $+0.0148$         & $+0.0505$          & $-0.0697$         \\[6pt]
  $\widetilde{E}=0.8$ & $\widetilde{J} \approx$       & $\ 3.199 \pi$        & $2.633 \pi$       & $2.766 \pi$        & $2.421 \pi$     \\
  \ \ \ "             & $\widetilde{J}/(2\pi)-E-0.5=$ & $\ +0.299$           & $+0.0166$         & $+0.0834$          & $-0.0895$         \\[6pt]
  $\widetilde{E}=0.9$ & $\widetilde{J} \approx$       & $\ 3.423 \pi$        & $2.832 \pi$       & $3.017 \pi$        & $2.604 \pi$    \\
  \ \ \ "             & $\widetilde{J}/(2\pi)-E-0.5=$ & $\ +0.312$           & $+0.0162$         & $+0.109$           & $-0.0978$        \\[6pt]
  $\widetilde{E}=1$   & $\widetilde{J} \approx$       & $\ 3.585 \pi$        & $3.029 \pi$       & $3.243 \pi$        & $2.811 \pi$    \\
  \ \ \ "             & $\widetilde{J}/(2\pi)-E-0.5=$ & $\ +0.293$           & $+0.0143$         & $+0.122$           & $-0.0946$         \\[6pt]
  $\widetilde{E}=1.1$ & $\widetilde{J} \approx$       & $\ 3.705 \pi$        & $3.223 \pi$       & $3.439 \pi$        & $3.038 \pi$    \\
  \ \ \ "             & $\widetilde{J}/(2\pi)-E-0.5=$ & $\ +0.253$           & $+0.0124$         & $+0.120$           & $-0.0812$         \\[6pt]
  $\widetilde{E}=1.2$ & $\widetilde{J} \approx$       & $\ 3.799 \pi$        & $3.417 \pi$       & $3.608 \pi$        & $3.279 \pi$    \\
  \ \ \ "             & $\widetilde{J}/(2\pi)-E-0.5=$ & $\ +0.200$           & $+0.00856$        & $+0.104$           & $-0.0607$         \\[6pt]
  $\widetilde{E}=1.3$ & $\widetilde{J} \approx$       & $\ 3.876 \pi$        & $3.611 \pi$       & $3.754\pi$         & $3.525 \pi$    \\
  \ \ \ "             & $\widetilde{J}/(2\pi)-E-0.5=$ & $\ +0.138$           & $+0.00544$        & $+0.0770$          & $-0.0374$         \\[6pt]
  $\widetilde{E}=1.4$ & $\widetilde{J} \approx$       & $\ 3.941 \pi$        & $3.805 \pi$       & $3.883 \pi$        & $3.768 \pi$    \\
  \ \ \ "             & $\widetilde{J}/(2\pi)-E-0.5=$ & $\ +0.0707$          & $+0.00252$        & $+0.0410$          & $-0.0161$         \\[6pt]
  $E=1.5$             & $J \approx$                   & $\ 4 \pi$            & $4 \pi$           & $4 \pi$            & $4 \pi$            \\
  \ \ \ "             & $|J/(2\pi)-E-0.5|<$           & $\ 2\times10^{-14}$  & $2\times10^{-15}$ &$9\times10^{-15}$   & $2\times10^{-15}$      \\[6pt]
  $\widetilde{E}=1.6$ & $\widetilde{J} \approx$       & $\ 4.055 \pi$        & $4.196 \pi$       & $4.110 \pi$        & $4.219 \pi$    \\
  \ \ \ "             & $\widetilde{J}/(2\pi)-E-0.5=$ & $\ -0.0724$          & $-0.00201$        & $-0.0450$          & $+0.00934\times10^{-3}$         \\[6pt] \hline

\end{tabular}
\end{center}
\end{table}

\begin{figure}[h]
\centering
\includegraphics[scale=0.38]{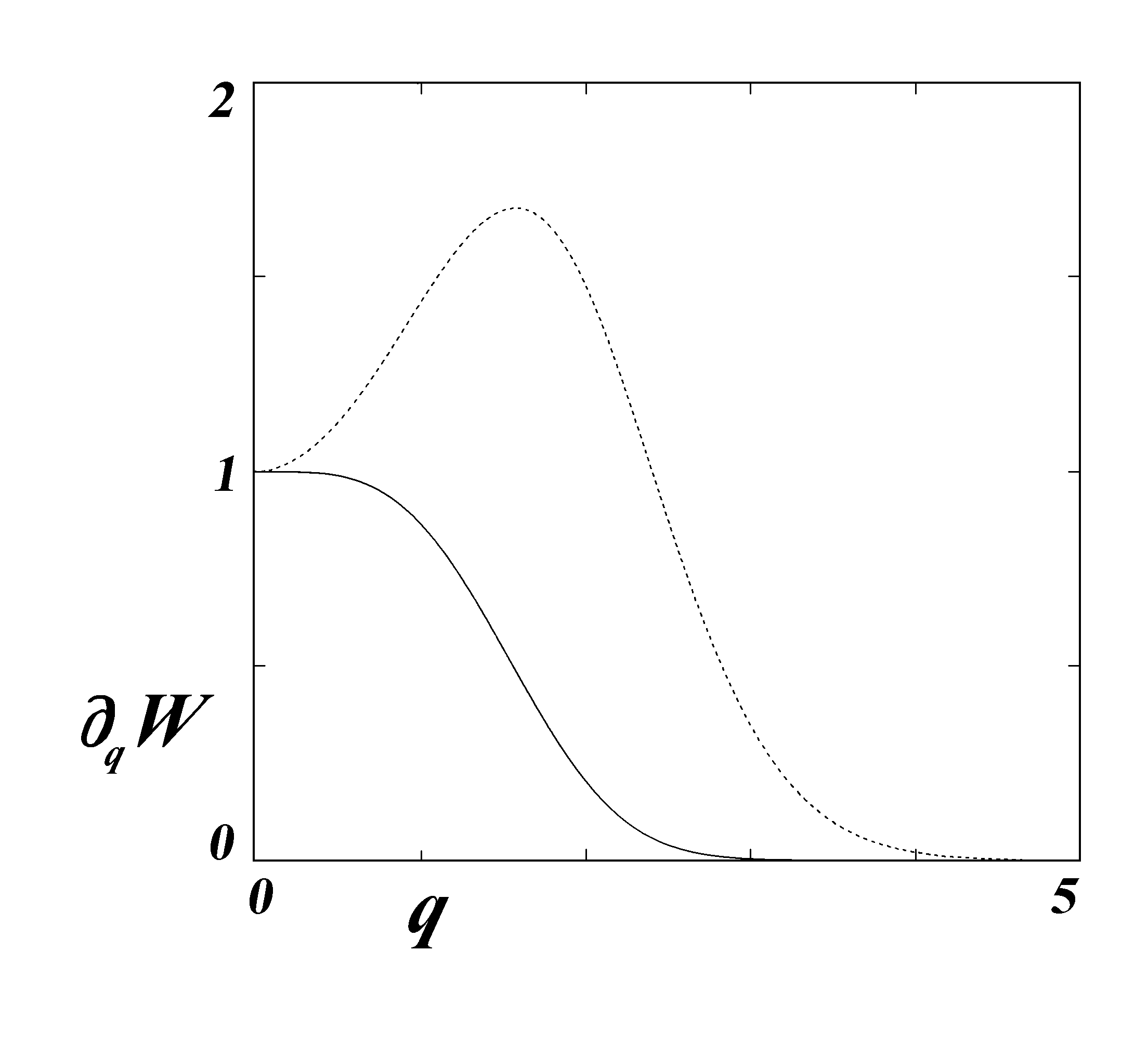}
\caption{\small Conjugate momentum for Case C as a function of $q$ for the ground state, $E=0.5 \hbar \omega$ or $E=0.5$ in natural units.  The solid line is the conjugate momentum.  The short-dash line is the conjugate momentum multiplied by the divergence factor $\exp(q^2/2)$.  Presentations are in natural units.}
\end{figure}

\begin{figure}[h]
\centering
\includegraphics[scale=0.38]{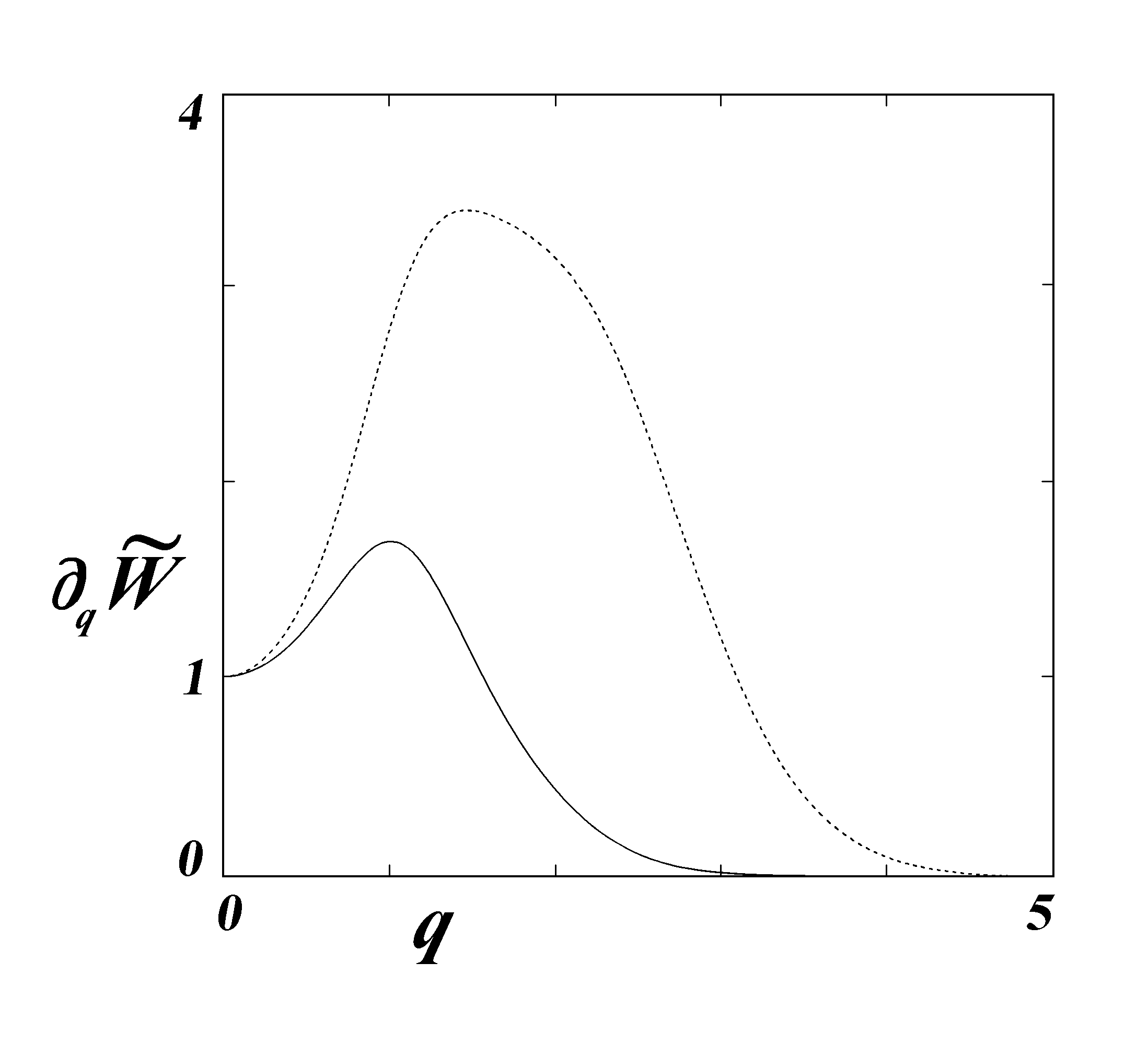}
\caption{\small Virtual conjugate momentum for Case C as a function of $q$ for the virtual state, $\widetilde{E}=\hbar \omega$ or $\widetilde{E}=1$ in natural units.  The solid line is the virtual conjugate momentum. The short-dash line is the conjugate momentum multiplied by the divergence factor $\exp(q^2/2)$.  Presentations are in natural units.}
\end{figure}

Every finite virtual energy $\widetilde{E}$ specifies a particular virtual action variable $\widetilde{J}$ as a function of the initial value,
$\partial W/\partial q|_{q=0}$, as exhibited by Table 1 (the value of $\partial W/\partial q|_{q=0}$ specifies the particular case in Table 1).  Table 1 exhibits Lipschitz continuity where  moderate changes in energy, $\Delta E=0.1$, make moderate, smooth changes $J$ or $\widetilde{J}$ that in turn imply, by Milne quantization, Lipschitz continuity with respect to energy.  This Lipschitz continuity also supports $\lim_{\widetilde{E} \to E_{\mbox{\tiny eigenvalue}}} \widetilde{W}(\widetilde{E},q) = W(E_{\mbox{\scriptsize eigenvalue}},q)$.  As earlier stated stated, the other initial values have been fixed, $\partial^2 W/\partial q^2 |_{q=0}=0$ and $W(0)=0$, for computational ease.  As the algorithm $J \approx 4W(\{W,\partial_q W,{\partial^2}_q W\}|_{q=0};E,10)$ where in turn $W(W(\{W,\partial_q W,{\partial^2}_q W\}|_{q=0};E,q)$ has been uniquely determined by the QSHJE, the numerical computed $J$ is uniquely specified within computational round-off errors. This satisfies Hadamard uniqueness for the algorithm.  Every virtual $\widetilde{E}$, as exhibited by the rows of Table 1, is associated with an ensemble (Cases A--D) of only virtual $\widetilde{J}$s.    For the converse relationship, Table 1 also implies from the columns for Cases A--D that any virtual $\widetilde{J}$ is associated with an ensemble of only virtual $\widetilde{E}$s.
For eigenvalues of $E$ and $J$, each $E$ specifies a unique $J$ and vice versa consistent with the relationship $J/(2\pi)=E+0.5$ in the quantum domain in natural units [classically $J/(2\pi)=E$] regardless of which particular set of initial values is used.  This is consistent with the observation that a bound-state quantum reduced action, $W(\{W,\partial_q W,{\partial^2}_q W\}|_{q=0};E,q)$, generates a particular quantum trajectory per its constants of quantum motion [\ref{bib:prd29}--\ref{bib:rc}]. Each particular trajectory manifests a unique microstate of a bound-state wave function.  A bound-state wave function $\psi$ is described by a real wave function to within an arbitrary phase shift [\ref{bib:prd29},\ref{bib:fpl9}].  Again as the QSHJE is a third-order differential equation, the quantum reduced action contains more information than the wave solution $\psi$ of the second-order SSE.  While knowledge of any single bound-state $W$ renders the bound-state $\psi$, the converse is generally not true [\ref{bib:prd29},\ref{bib:fpl9},\ref{bib:rc},\ref{bib:fm}]. Full knowledge of a bound-state $\psi$ is insufficient to specify a unique bound-state $W$.

The columns of Table 1 show that the numerically computed $\widetilde{J}$ vary smoothly with $\widetilde{E}$ for any Case.  Table 2 examines the behavior of the
calculated $\widetilde{J}$ near the ground state with finer-spaced $\widetilde{E}$ for every Case.  The behavior of $\widetilde{J}-2\pi$ in Table 2 and of
$\widetilde{J}$ in Table 1 suggests that for Cases A, C, and D the function, $\widetilde{J}_i(\widetilde{E}),\ \mbox{Case }i=A,C,D$ would trace smooth
case-dependent curves in the $(J\cup\widetilde{J},E\cup\widetilde{E})$-plane as a function of $\widetilde{E}$.  As the domains of $\widetilde{J}$ and $\widetilde{E}$ are segmented by their respective limit points, the $J$s and $E$s, Jacobi's theorem is substantiated at $q=\infty$ for bound states. The monotonic behavior of $W(E,q)$ with $q$ [\ref{bib:fm}] supports Jacobi's theorem for finite $q$.   The values for $\widetilde{J}-2\pi$ for $\widetilde{E}=0.499,0.501$ for any particular case on Table 2 only vary in the fourth significant place, which supports Lipschitz continuity for $\Delta E = 0.001$ and subsequently  a numerical Jacobi's theorem by finite differences. A comparison of Tables 1 and 2 supports the realization of Jacobi's theorem for bound states where $\lim_{\widetilde{E} \to E} \partial_{\widetilde{E}} \widetilde{W} = \partial_E W$.  The extended curves for all functions $\widetilde{J}_i(\widetilde{E})$ and $J_i(E)$ on the $(J\cup\widetilde{J},E\cup\widetilde{E})$-plane intersect the limit points $(J,E)$ where $J/(2\pi)=E+1/2$ in natural units. These well-behaved intersections support numerical shooting recipes, which are discussed in the next paragraph [\ref{bib:st}].  It follows that the curves  $\widetilde{J}_i(\widetilde{E})$ form segments that are bounded by the limit points $J=2\pi[\widetilde{J}/(2\pi)]<\widetilde{J}<J+2\pi$ and $E=[\widetilde{E}+0.5]-0.5<\widetilde{E}<E+1$ in natural units where $[\chi]$ is here the
piecewise-defined greatest integer function that returns the greatest integer less than or equal to $\chi$.

While the linear expression $J/(2\pi)=E+1/2$ in natural units holds for eigenvalues of the linear harmonic oscillator, it does not hold for non-eigenvalues $\widetilde{E}$ and $\widetilde{J}$ as shown in the prior paragraph and by Tables 1 and 2.  A numerical shooting recipe for finding the eigenvalues $J$s and $E$s may be used here and also for any other potential including ones without any symmetry [\ref{bib:st}].  Its first step computes $\widetilde{J}=2(\widetilde{W}|_{q=+\infty}-\widetilde{W}|_{q=-\infty})$ with a trial energy $\widetilde{E}$.  The next step observes the computed $\widetilde{J}$, then corrects with a second trial energy, and repeats the recipe until the calculated action variable is within a specified closeness to Milne quantization for the action variable [\ref{bib:pr35}]. If the dependence of $\widetilde{W}(\{W,\partial_q W,{\partial^2}_q W\}|_{q=q_0};\widetilde{E},\pm \infty)$ upon $\widetilde{E}$ were known in closed form, then the Newton-Raphson method could be used to estimate the second and succeeding trial energies.  Regardless of whether $\widetilde{W}$'s dependence upon $\widetilde{E}$ is known in closed form, the third and succeeding trial energies may be always calculated by {\itshape regula falsi} or Lagrangian interpolation.  {\itshape Regula falsi} is general, convergent and needs neither Bornian probability nor knowledge of $\psi$.  Its rate of convergence indicates how well posed the quantum reduced action is.  Again, an innately monotonic $W$ is better behaved than $\psi$.

\begin{table}
\begin{center}
\caption{\small Numerically calculated values of $\widetilde{J}-2\pi \hbar$ or $\widetilde{J}-2\pi$ in natural units for non-eigenvalue energies in the vicinity near the ground state of the linear harmonic oscillator are presented in natural units. Numerical calculated values of $J-2\pi$ in natural units for the ground state ($E=0.5$) of the linear harmonic oscillator are also presented. Calculations are done for Cases A through D.  Presentations of values are in natural units.}

\begin{tabular}{r|cccc}
\hline
                                      &                         &                         &                         &                   \\[-6pt]
     Case:                            &  \  A                   &  \ B                    &    \ C                  & \  D              \\
$(\partial W/\partial q)|_{q=0}$      &    $\ 0.5$              & $\ (2E)^{1/2}$          & $\ 1$                   &  $\ 2$      \\[3pt] \hline
                                      &                         &                         &                         &                   \\[-6pt]
$\widetilde{E}=0.499$                 & $-0.004510\pi$          & $-0.002257\pi$          & $-0.002255\pi$          & $-0.001128\pi$     \\[3pt]
$E=0.5\ \ \ $                         &$+5.329\pi\times10^{-15}$&$+3.997\pi\times10^{-15}$&$+3.997\pi\times10^{-15}$&$1.312\pi\times10^{-15}$\\[3pt]
$\widetilde{E}=0.501$                 & $+0.004517\pi$          & $+0.002256\pi$          & $+0.002258\pi$          & $+0.001129\pi$    \\[6pt] \hline
\end{tabular}
\end{center}
\end{table}

A quantum system with increasing energy, whether described by eigenvalue or even virtual energies, approaches the analogous classical system consistent with the Bohr correspondence principle [\ref{bib:tl}] with a caveat for the linear harmonic oscillator for the $n^{\mbox{\scriptsize th}}$ eigenvalue that the approach is displaced by the quantum term $\pi = J - J_{\mbox{\scriptsize classical}} = 2n\pi - (2n-1)\pi$ no matter how great $n$ becomes.    This is illustrated for the linear harmonic oscillator by considering the value of $\widetilde{J}/(2\pi) -\widetilde{E}-0.5$ for the virtual energies $\widetilde{E}=1,2,4,8,16,32$, which lie midway between neighboring eigenvalue energies.  The results are presented in Table 3 for Case B  where the initial conjugate momentum at $q=0$ is given by $\partial_q \widetilde{W} = (2E)^{1/2}$ in natural units.  While Case B is a representative case, the value of  $\widetilde{J}/(2\pi) -\widetilde{E}-0.5$ is microstate dependent.  Note the decrease on Table 3 in the absolute value of $|\widetilde{J}/(2\pi) -\widetilde{E}-0.5|$ with increasing $\widetilde{E}=1,2,4,8,16,32$. The behavior of the value $\widetilde{J}/(2\pi) -\widetilde{E}-0.5$ with increasing $\widetilde{E}$  approaches that for neighboring eigenvalues.  This when combined with the results for Case B on Table 1 suggest that $\widetilde{J}$ be Lipschitz continuous with regard to energy.  Subsequently, this too supports the validity of Jacobi's theorem for time parametrization for bound states.  (In calculation for the value $\widetilde{J}/(2\pi) - \widetilde{E}  - 0.5$ for $\widetilde{E}=32$, the corresponding classical turning point is at $q=8$. As a numerical precaution, Rkadapt computations were extended to $q=11.5$ using 4600 points to drive the conjugate momentum to Rkadapt's effective zero, which due to accumulated rounding errors is at the fifteenth significant figure, at $q=11.47$ in natural units.)

\begin{table}
\begin{center}
\caption{\small The deviation from zero for the value of $\widetilde{J}/(2\pi) - \widetilde{E} -0.5$ for various integer $\widetilde{E}$s in natural units for the linear harmonic oscillator.  Numerically calculated values of $\widetilde{J}/(2\pi) - \widetilde{E} -0.5$ in natural units are presented for various integer $\widetilde{E}$s for Case B where $\partial_q \widetilde{W} = (2\widetilde{E})^{1/2}$ in natural units.  As integer $\widetilde{E}$ increases, $|\widetilde{J}/(2\pi) - \widetilde{E} -0.5|$ in natural units decreases consistent with the Bohr correspondence principle.  Presentations of values are in natural units.}

\begin{tabular}{c|cccccc}
\hline
                                          &                  &                 &                 &                      &                 &\\[-6pt]
     $\widetilde{E}$                      &   \ 1            &    \ 2          &   \ 4           &   \ 8                &    \ 16           &   \ 32  \\[3pt] \hline
                                          &                  &                 &                 &                      &                \\[-6pt]
$\widetilde{J}/(2\pi)-\widetilde{E}-0.5$\ & $\ \ +0.0287\ \ $          & $\ \ -0.00883\ \ $      & $\ -0.00240\ $      &$-6.16\times10^{-4}$&$-1.55\times10^{-4}$&$-3.89\times10^{-5}$\\[6pt] \hline

\end{tabular}
\end{center}
\end{table}

For completeness, quantization of the action variable for the ground state of the linear harmonic oscillator for asymmetric initial values was initially
examined in unpublished Ref.\ \ref{bib:prvda25}.  Here, an asymmetric conjugate momentum, which manifests an asymmetric microstate of the ground state $\psi$,
is examined in more detail.   The other constants of the motion (initial values) for the asymmetric quantum reduced action are given by $W(0)=0,\ \partial_q W|_{q=0}=1$ and $\partial_q^2 W|_{q=0}=0.5$.  The conjugate momentum is exhibited on Figure 3. The computed action variable, $J$ in natural units is given by Rkadapt as

\begin{equation}
J \approx\ 2[ W(10)-W(-10)]\ =\ 2( \underbrace{1.815\,774\,989\,921\,758}_{+W(10)} \ \underbrace{+\ 1.325\,817\,663\,668\,034}_{-W(-10)} )\ =\ 2\pi
\label{eq:wasm}
\end{equation}

\noindent The numerical accuracy of Rkadapt, fifteen significant figures, for this asymmetric case is exhibited by $J/(2\pi)-E-0.5=1.11 \times 10^{-15}$.  This example is examined further in \S3.2.

\begin{figure}[h]
\centering
\includegraphics[scale=0.38]{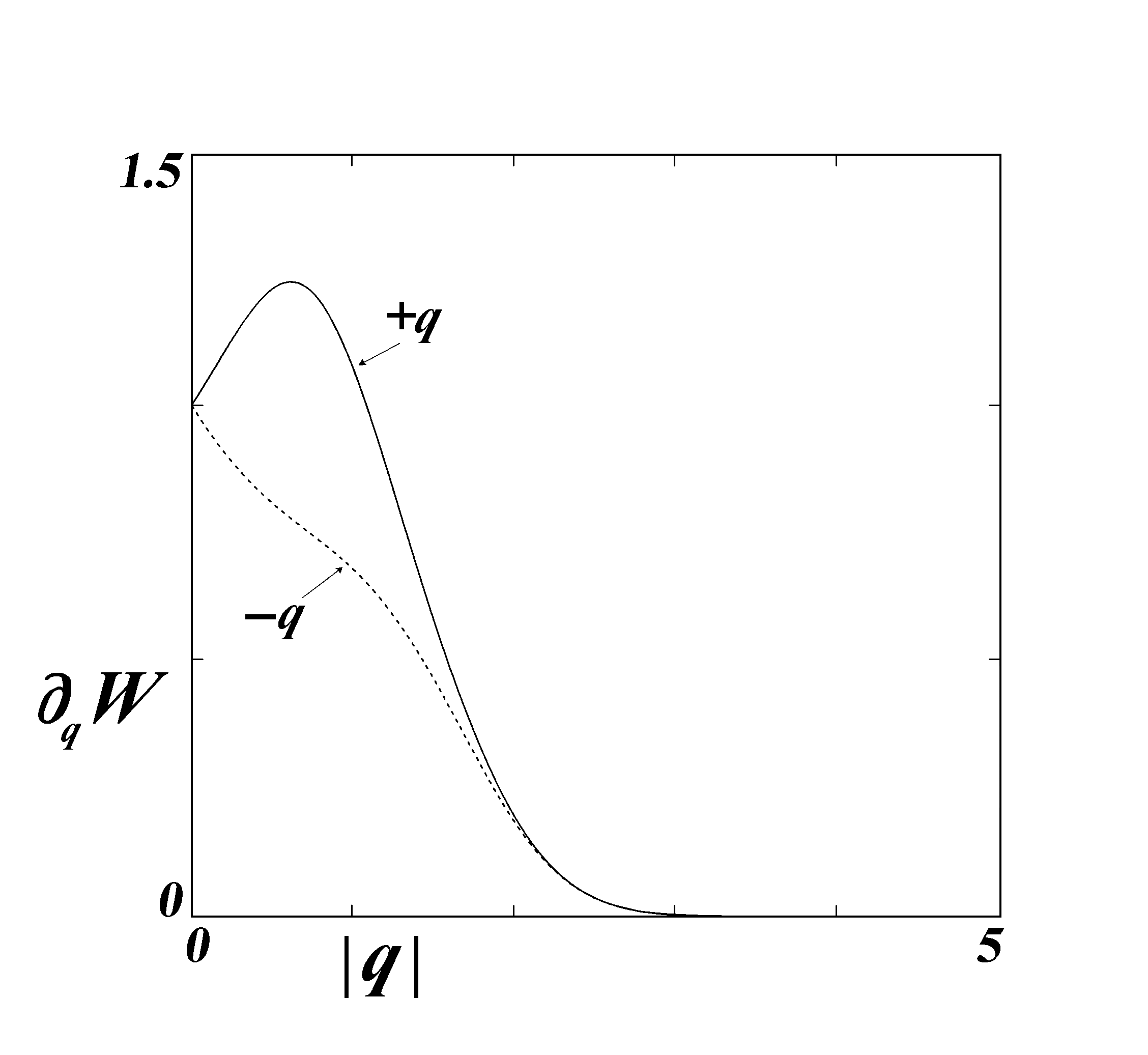}
\caption{\small Asymmetric conjugate momentum in phase space for the ground state, $E=0.5$ in natural units. The set of initial values for the asymmetric reduced action are
$\{W,\partial_q W,{\partial^2}_p W\}|_{q=0} = \{0,1,0.5\}|_{q=o}$ in natural units.}
\end{figure}

\subsection{Numerical Time Parametrization}

An approximate time parametrization for the linear harmonic oscillator may be generated by Jacobi's theorem using finite differences by reversing $\lim_{\widetilde{E} \to E} \partial_{\widetilde{E}} \widetilde{W} = \partial_E W$, where

\begin{eqnarray}
t-\tau &=& \frac{\partial W(\{W,\partial_q W,{\partial^2}_q W\}|_{q=0};E,q)}{\partial E} \nonumber \\
       &=& \lim_{\epsilon \to 0} \frac{\widetilde{W}(\{W,\partial_q W,{\partial^2}_q W\}|_{q=0};\widetilde{E+\epsilon},q)-\widetilde{W}(\{W,\partial_q W,{\partial^2}_ q W\}|_{q=0};\widetilde{E-\epsilon},q)}         {2\epsilon} \nonumber \\
     &\approx& \frac{\widetilde{W}(\{W_E,\partial_q W_E,{\partial^2}_q W_E\}|_{q=0};\widetilde{E+\epsilon},q)-\widetilde{W}(\{W_E,\partial_q W_E, {\partial^2}_q W_E|\}|_{q=0};\widetilde{E-\epsilon},q)}{2\epsilon}
               \bigg|_{0<\epsilon \ll 1}
\label{eq:njt}
\end{eqnarray}

\noindent  where $W_E \equiv  W(\{W,\partial_q W,{\partial^2}_q W\}|_{q=0};E,q)$, as discussed in the next paragraph.  The third line of Eq.\ (\ref{eq:njt}) shows Jacobi's principle executed numerically by a finite difference method which assumes Lipschitz continuity. Conversely, broad application of Jacobi's principle by finite deference methods implies Lipschitz continuity. The virtual reduced actions, $\widetilde{W}$s in the second and third lines of Eq.\ (\ref{eq:njt}) have virtual energies $\widetilde{E\pm\epsilon}$ located in the different virtual energy domains that are separated by their common limit point $E$.  Explicitly, a virtual $\widetilde{E \pm \epsilon}$ is displaced from the common limit point $E$ by $\pm \epsilon$.   The quantum trajectory and time parametrization are dependent upon the particular microstate as specified by the initial values of $W|_{q=q_0}$ on the first line of the right side of Eq.\ (\ref{eq:njt}).  This satisfies Hadamard for uniqueness of time parametrization.

We take the precaution that the numerical process to parametrize time is consistent with the variational principles of the associated underlying Lagrangian.  The finite difference algorithm inherently samples $W$ at different energies, here at $\widetilde{E \pm \epsilon}$, to implement Jacobi's theorem. The explicit values for the initial values at $q=q_0$ for each sampled $W$ in Eq.\ (\ref{eq:njt}) must be compatible with the variational principle. Underlying Hamilton's principal function, $S=W-Et$, is the Lagrangian, $L=dS/dt$.  A Lagrangian of the form $L(q,\dot{q},\ddot{q},t)$ under Hamilton's principle will have nil variation of $q,\ \dot{q}$, and $\ddot{q}$ at its end points of its trajectory in the $q,t$-plane. As the QSHJE is third order, the initial values $\{W,\partial_q W,{\partial^2}_qW\}|_{q=q_0}$ of each sample $\widetilde{W}$ are held fast to satisfy Hamilton's principle for the finite difference approximation in Eq.\ (\ref{eq:njt}).  These common initial values specify a common analogous microstate for the sampled quantum reduced actions.  Again, Faraggi and Matone have developed the relationships between $\{\dot{q},\ddot{q}\}$ and $\{\partial_q W,{\partial^2}_q W\}$ in Ref.\ \ref{bib:fm}.    The initial values $\{W,\partial_q W,{\partial^2}_qW\}|_{q=q_0}$ of the quantum reduced actions for displaced energy, $\widetilde{E \pm \epsilon}$, must be the same as those initial values, $\{W,\partial_q W,{\partial^2}_qW\}|_{q=q_0}$, for the un-displaced quantum reduced action with energy $E$ in Eq. (\ref{eq:njt}).  At the other end point $q=\infty$, the quantum trajectory in the $q,t$-plane for bounds states asymptotically levels off to its terminal value as exhibited by Figure 4.  This is consistent with the quantum reduced action, which has an asymptotic stable nodal singularity at $q\to\infty$ [\ref{bib:prd25}], to level off asymptotically towards its terminal value, $J/4$, consistent with the quantum conjugate momentum asymptotically reducing to zero as exhibited by Figures 1--3 and later confirmed in \S3 by Eqs.\ (\ref{eq:gcm}) and (\ref{eq:gcm2}).  This asymptotic behavior of $W$ and $\widetilde{W}$ at its stable nodal singularity at $q = \pm  \infty$  substantiates that $W$ and $\widetilde{W}$ are not constants in the finite $q$-region, $-\infty \le q \le +\infty$.  The asymptotic behavior of $W$ and $\widetilde{W}$ maintains ${\cal{C}}^2$ continuity throughout the asymptotic region.  This is consistent with the QEP's requirement that $W$ never be a constant.

Using Case B for time parametrization is not permitted. Case B changes microstate specification even for small variations of $E$ that would, for time parametrization, be incompatible with the prescribed initial values for $L(q,\dot{q},\ddot{q},t)$.   In \S3.4 for completeness, Jacobi's theorem is implemented analytically and simply without resorting to numerical finite differences for the quantum square well where the quantum reduced action is known in familiar terms of linear, trigonometric, and hyperbolic functions.

Sample time parametrizations have been calculated using Eq.\ (\ref{eq:njt}) with $\epsilon = 10^{-5}$ in natural units.  Time parametrizations for the ground state ($E=0.5$ in natural units) and the virtual state ($\widetilde{E}=1$) are graphically presented respectively by Figures 4 and 5 for microstates specified by Cases A, C, and D.  The quantum trajectories in the $q,t$-plane are dependent upon the particular Case (particular microstate).  The dispersion of the quantum trajectories, as shown by Figure 4 for the ground state, is another manifestation of microstates. Figure 5 exhibits the trajectories for $E=0.5,1,1.5$ in natural units representing the ground state, a virtual state, and the first excited state respectively.  The initial values at $q=0$ are $\{W,\partial_q W,{\partial^2}_q W\}|_{q=0} = \{0,1,0\}|_{q=0}$ in natural units for all quantum trajectories on Figure 5.

\begin{figure}[h]
\centering
\includegraphics[scale=0.38]{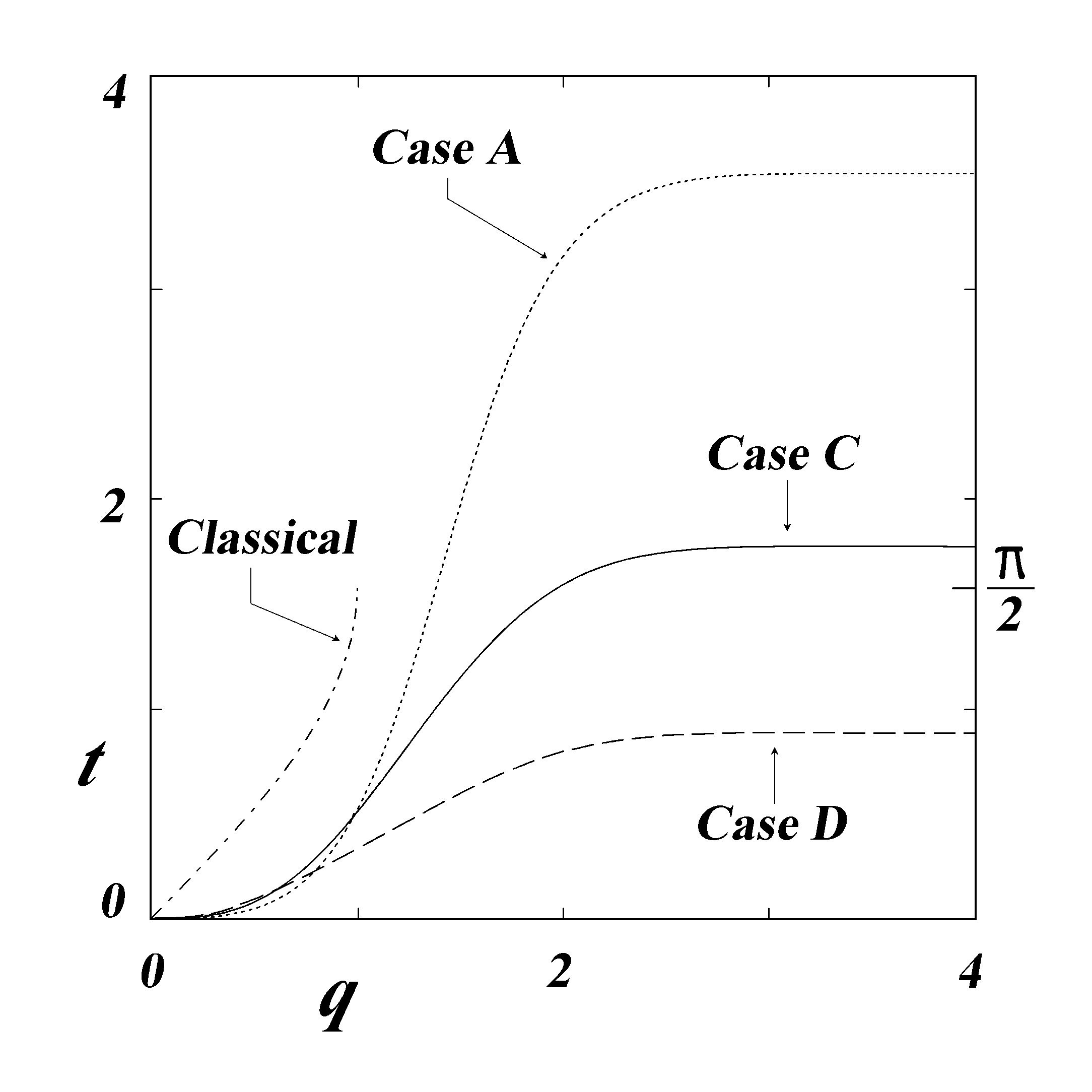}
\caption{\small Time, $t$ as a function of $q$ for microstates Cases A, C and D for the ground state and also for the classical linear harmonic oscillator with $E=0.5$
in natural units.  The transit time for the classical oscillator for a quarter-cycle of oscillation is $\pi/2$ which is presented for reference.}
\end{figure}

\begin{figure}[h]
\centering
\includegraphics[scale=0.38]{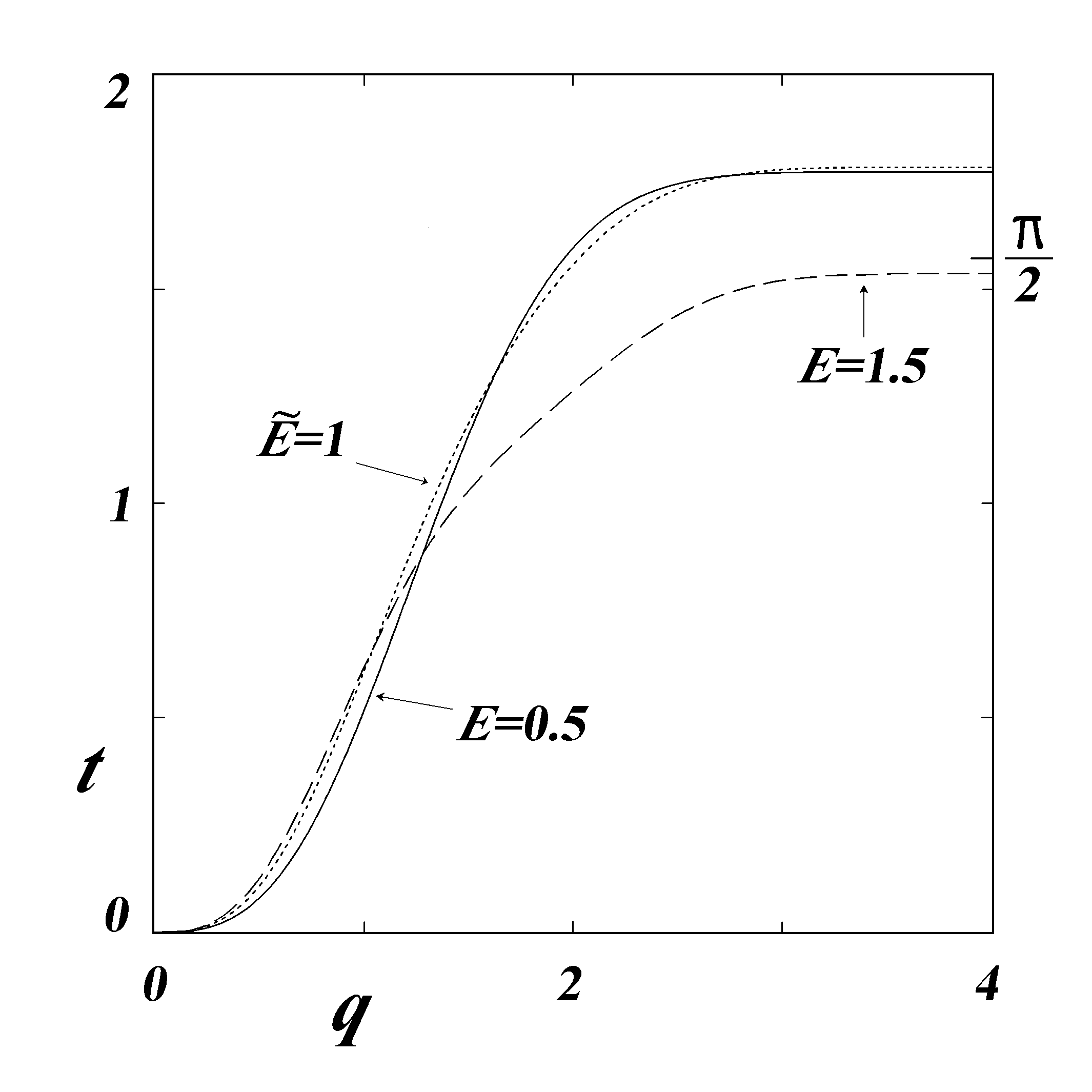}
\caption{\small Time, $t$ as a function of $q$ for microstate Case C for the ground state, $E=0.5$, for the non-eigenvalue energy, $\widetilde{E}=1$, and
for the first excited state, $E=1.5$ (all in natural units).  The transit time for the classical oscillator for a quarter-cycle of oscillation is $\pi/2$ which is presented for reference.}
\end{figure}

\subsection{Cycle Time}

A prime characteristic of the linear harmonic oscillator of classical mechanics is its constant cycle time $T_{\mbox{\scriptsize classical}}$, the duration to complete one cycle of $q$, independent of the amplitude of the sinusoidal motion. Numerically, $T$ may be evaluated by $T = \oint \partial_E W\, dq$ circumscribing its branch cut in complex plane between the asymptotic stable nodal singularities of $W$. A brief recital of the classical linear harmonic oscillator is presented.  Its frequency, $\nu_{\mbox{\scriptsize classical}}$ is given by

\begin{equation}
\frac{\partial H_{\mbox{\scriptsize classical}}}{\partial J_{\mbox{\scriptsize classical}}} = \nu_{\mbox{\scriptsize classical}} = const
\label{eq:nu}
\end{equation}

\noindent  where H is the Hamiltonian.  In classical mechanics, $\nu_{\mbox{\scriptsize classical}} = \omega/(2\pi) = (2\pi)^{-1}$ in natural units.  The duration to complete one cycle is $T_{\mbox{\scriptsize classical}}= \nu^{-1} = 2\pi/\omega = 2\pi$ in natural units.  Both $\nu_{\mbox{\scriptsize classical}}$ and $T_{\mbox{\scriptsize classical}}$ are independent of energy or sinusoidal amplitude for the classical linear harmonic oscillator.

For the quantum linear harmonic oscillator, the quantum term $-\hbar^2 \langle W;q \rangle/(4m)$ on the right side of the QSHJE, Eq.\ (\ref{eq:qshje}),
effects the frequency, $\nu$, so that it is no longer a constant independent of energy. The quantum effects may be investigated by studying the difference between either $T/4$ or $\widetilde{T}/4$ and $T_{\mbox{\scriptsize classical}}/4=\pi/2$ where symmetry permits us to investigate the transit time for only a quarter-cycle (one quarter of the period of oscillation or the transit time $t|^{\infty}_{q=0}$).  We present these differences of periodicity between the quantum and classical linear harmonic oscillators in Table 4.  We investigate the differences for eigenvalue $E = 0.5,1.5,2.5,3.5,4.5,9.5,14.5$ interspersed with integer $\widetilde{E}=1,2,3,4,5,10,15$.  Case C, whose quantum reduced action at $q=0$ has already been shown to osculate with the corresponding classical reduced action, inspires the choice of which set of initial values to use to exemplify solving the QSHJE.  All initial values at $q=0$ for (virtual) quantum reduced action are given by $W,\widetilde{W}=0$ and for its second derivative by ${\partial^2}_q W, {\partial^2}_q \widetilde{W} =0$. The initial values for (virtual) conjugate momentum are presented in Table 4 for various (virtual) energies.
The QSHJE, Eq.\ (\ref{eq:qshje}), is solved by Rkadapt as described before.  The quarter-cycle transit time between $q=0$ and $q=10$ is numerically calculated by Jacobi's theorem, Eq.\  (\ref{eq:njt}).  The corresponding transit time for the classical linear harmonic oscillator
is a constant $\pi/2$ in natural units regardless of amplitude.   Table 4 presents the differences between classical and quantum transit times, $T/4-\pi/2$ and
$\widetilde{T}/4-\pi/2$ in natural units.  Table 4 also exhibits that the quantum transit times with increasing energy quickly approach the constant classical transit time consistent with the Bohr correspondence principle even for deeply non-eigenvalues of integer $\widetilde{E}$, albeit not as fast as for eigenvalue $E$.  While the transit times for integer $\widetilde{E}$ are virtual, they are not ``quantum spooky".  Also note that on Table 4 the values of $T/4-\pi/2$ are positive for symmetric eigenfunction $\psi$s where $E=0.5,2.5,4.5,\cdots$ and negative for antisymmetric  $\psi$s where $E = 1.5,3.5,5.5,\cdots$.  Other microstates, which have different initial values and different physics, would render different transit times for a quarter-cycle, for they also have different trajectories as implied by Figures 3 and 4.

If the quantum equivalence of Eq.\ (\ref{eq:nu}), with eigenvalue $E=H_{\mbox{\scriptsize quantum}}$, were prosecuted numerically with sample points spaced at the eigenvalues of $E$ and $J$, as suggested by Faraggi and Matone [\ref{bib:fm3}] for semi-classical results, it would result in aliasing where the effects of microstates inherent to $\psi$ and $W$ would be missed.  All microstates on Table 1 for any eigenvalue $E$ exhibit that all microstates are Lipschitz continuous for $\Delta \widetilde{E}=0.1$.

\begin{table}
\begin{center}
\caption{\small The calculated difference between quantum and classical transit times over a quarter-cycle for the linear harmonic oscillator.  The classical transit
time is $T_{\mbox{\scriptsize classical}}=\pi/2$ in natural units.  The value of initial conjugate momentum for each energy is specified in the Table.  Otherwise,
all the other initial values for $W$ or $\widetilde{W}$ and ${\partial^2}_q W$ or ${\partial^2}_q \widetilde{W}$ at $q=0$ are nil.  The differences between quantum
and classical transit times are tabled in natural units for various eigenvalue energies, $E$, as $T/4-\pi/2$ and for various non-eigenvalue energies,
$\widetilde{E}$, as $\widetilde{T}/4-\pi/2$.  The convergence of $T/4$ or $\widetilde{T}/4$ to $\pi/2$ with increasing energy is consistent with the Bohr correspondence principle.}

\begin{tabular}{c|c|c|c|c|c|c}
\hline
       &                    &                        & \hspace{.75in} &                 &                                &                         \\ [-6pt]
  $E$  &$\partial_qW|_{q=0}$&    $T/4-\pi/2$         &                & $\widetilde{E}$ &$\partial_q\widetilde{W}|_{q=0}$& $\widetilde{T}/4-\pi/2$ \\ [3pt] \cline{1-3} \cline{5-7}
       &                    &                        &                &                 &                                &                         \\ [-6pt]
  0.5  &     1              & $+2.02 \times 10^{-1}$ &                &     1           & $2^{1/2}$                      &  $+2.13 \times 10^{-1}$      \\
  1.5  &   $3^{1/2}$        & $-3.58 \times 10^{-2}$ &                &     2           &  2                             & $-1.19 \times 10^{-1}$      \\
  2.5  &   $5^{1/2}$        & $+1.45 \times 10^{-2}$ &                &     3           & $6^{1/2}$                      & $+8.12 \times 10^{-2}$      \\
  3.5  &   $7^{1/2}$        & $-7.64 \times 10^{-3}$ &                &     4           & $2^{5/2}$                      & $-6.14 \times 10^{-2}$      \\
  4.5  &     3              & $+4.72 \times 10^{-3}$ &                &     5           & $10^{1/2}$                     & $+5.05 \times 10^{-2}$      \\ [6pt]
  9.5  &   $19^{1/2}$       & $-1.08 \times 10^{-3}$ &                &    10           & $2 \times 5^{1/2}$             & $-2.50 \times 10^{-2}$      \\
 14.5  &   $29^{1/2}$       & $+4.66 \times 10^{-4}$ &                &    15           & $30^{1/2}$                     & $+1.66 \times 10^{-2}$      \\ [3pt] \hline

\end{tabular}
\end{center}
\end{table}

\clearpage

\section{Finite Square Well}

Numerical results of \S 2 support time parametrization by Jacobi's theorem for bound states, albeit the time parametrization is dependent upon
the particular microstate (quantum trajectory).  Closed-form analyses are presented in this section.  Finite difference algorithms are not used in the investigation of time parametrization for the finite square well.  Quantum reduced actions, conjugate momenta, and quantum trajectories for finite square wells may be expressed in closed form in familiar terms of trigonometric, inverse trigonometric, exponential, and hyperbolic functions.  Quantum reduced actions are developed for the bound states with eigenvalue energies and for virtual states with non-eigenvalue energies to substantiate using Jacobi's theorem for quantized energies and action variables.

\subsection{General Solution of QSHJE}

A general solution for the quantum reduced action $W$ of the phenomenological QSHJE, Eq.\ (\ref{eq:qshje}), is given within an arbitrary integration constant by
[\ref{bib:eh}--\ref{bib:hm}]

\begin{equation}
W(E,q) = \hbar \arctan\left(\frac{A \phi(q) + B \vartheta(q)}{C \phi(q) + D \vartheta(q)}\right)
\label{eq:gw}
\end{equation}

\noindent where $\{\phi,\vartheta\}$ is a set of independent solutions of the associated phenomenological one-dimensional SSE,

\begin{equation}
-\frac{\hbar^2}{2m} \frac{\partial^2 \psi}{\partial q^2} + (V - E) \psi = 0.
\label{eq:se}
\end{equation}

\noindent As $\phi$ and $\vartheta$ are continuous with regard to $E$ even though, perhaps, not  eigenfunction,  $W(E,q)$ is Lipschitz continuous with regard to E.  The argument  of the arc tangent in Eq.\ (\ref{eq:gw}) is a bilinear transformation or M\"{o}bius transformation $\mathfrak{w}(q)$ that may be given by

\begin{equation}
\mathfrak{w}(q) = \tan \left(\frac{W(E,q)}{\hbar} \right) = \frac{A \phi(q) + B \vartheta(q)}{C \phi(q) + D \vartheta(q)}
\label{eq:frakw}
\end{equation}

\noindent for eigenvalue $E$ or $J$ and also for virtual $\widetilde{\mathfrak{w}}$.  For explicit eigenfunctions of $W$ or $\psi$ of the bound-state SSE, Faraggi and Matone have defined the ratio $w \equiv \psi^D/\psi$ where $\{\psi,\psi^D\}$ is explicitly the set of independent solutions with $\psi$ bound (convergent) and $\psi^D$ divergent [\ref{bib:fm}--\ref{bib:aef13}].  Then, the relationship expressed in QEP form between $\mathfrak{w}(q)$ and $W(E,q)$ for and only for eigenvalues of $E$ or $J$ is given by

\begin{eqnarray}
\mathfrak{w}(q)  =  \frac{A \phi(q) + B \vartheta(q)}{C \phi(q) + D \vartheta(q)} =  \frac{\mathfrak{A}\psi^D (q) + \mathfrak{B} \psi (q)}{\mathfrak{C} \psi^D (q) + \mathfrak{D} {\phi(q)}} & = & \frac{\mathfrak{A} \frac{\psi^D (q)}{\psi (q)} + \mathfrak{B}}{\mathfrak{C} \frac{\psi^D (q)}{\psi (q)} + \mathfrak{D}}  = \frac{\mathfrak{A} w(q) + \mathfrak{B}}{\mathfrak{C} w(q) + \mathfrak{D}} \label{eq:fmw} \\
                      & = & \frac{\mathfrak{A} + \mathfrak{B} \frac{\psi (q)}{\psi^D (q)} }{\mathfrak{C} + \mathfrak{D} \frac{\psi (q)}{\psi^D (q)}}  = \frac{\mathfrak{A}  + \mathfrak{B} w^{-1}}{\mathfrak{C} + \mathfrak{D} w^{-1}}
\label{eq:fmwi}
\end{eqnarray}

\noindent where by superposition $A \phi(q) + B \vartheta(q) = \mathfrak{A}\psi^D (q) + \mathfrak{B} \psi (q)$ and $C \phi(q) + D \vartheta(q) = \mathfrak{C} \psi^D (q) + \mathfrak{D} {\psi(q)}$.  If $\{\mathfrak{A},\mathfrak{B},\mathfrak{C},\mathfrak{D}\}=\{1,0,0,1\}$, then $\mathfrak{w}(q)=W(E,q)$, and if $\{\mathfrak{A},\mathfrak{B},\mathfrak{C},\mathfrak{D}\}=\{0,1,1,0\}$, then $\mathfrak{w}(q)=w^{-1}(q)$.  Equations (\ref{eq:fmw}) and (\ref{eq:fmwi}) exhibit the  inversion symmetry of the M\"{o}bius transform.

The coefficients forming the set $\{A,B,C,D\}$  in Eq.\ (\ref{eq:gw}) are real constants here, but $\{B,D\}$ may be complex for QEP [\ref{bib:fm}]. As the QSHJE is a third-order differential equation, the coefficients $\{A,B,C,D\}$ specifying the particular solution for $W$ for a given $E$ may be determined by the initial values for $\{W,\partial_q W,{\partial^2}_q W\}$ at some initial point $q_0$ with the normalization [\ref{bib:eh}]

\begin{equation}
AD-BC=1.
\label{eq:coefnorm}
\end{equation}

\noindent This normalization for the quantum trajectory representation replaces the Bornian (quantum) probability normalization for the $\psi$-representation.
One of the coefficients may be expressed in terms of the other three by Eq.\ (\ref{eq:coefnorm}).  Such has been tacitly done elsewhere where $W$ is specified by only three coefficients [\ref{bib:prd34},\ref{bib:fpl9}].

The conjugate momentum, $\partial_q W$, is also a solution to the phenomenological QSHJE, Eq.\ (\ref{eq:qshje}), as $W$ does not explicitly appear in the QSHJE.  The general
form of the conjugate momentum is represented consistent with Eqs.\ (\ref{eq:gw}) and (\ref{eq:coefnorm}) as

\begin{equation}
\frac{\partial W}{\partial q} =  \frac{\hbar (\overbrace{AD-BC}^1)\, \mathcal{W}(\phi,\vartheta)}{\underbrace{(A^2+C^2)\phi^2 + 2(AB+CD)\phi \vartheta +(B^2+D^2)\vartheta^2}_{(A\phi + B\vartheta)^2 + (C\phi + D\vartheta)^2 > 0}}
                              =  \frac{\hbar \mathcal{W}(\phi,\vartheta)}{(A^2+C^2)\phi^2 + 2(AB+CD)\phi \vartheta +(B^2+D^2)\vartheta^2}
\label{eq:gcm}
\end{equation}

\noindent where $\mathcal{W}(\phi,\vartheta)$ is the Wronskian of the independent solutions of the associated SSE given by
$\mathcal{W}(\phi,\vartheta) = (\partial_q \phi) \vartheta - \phi(\partial_q \vartheta)$.  As the phenomenological QSHJE is well posed, the conjugate momentum has Lipschitz continuity.  Microstate dependence of $\partial_q W$ is exhibited in Eq.\ (\ref{eq:gcm}) by the coefficients $\{A,B,C,D\}$.  As the SSE is a form of the Helmholtz equation, $\mathcal{W}(\phi,\vartheta) = const$, and the Wronskian is conserved in Eq.\ (\ref{eq:gcm}).   Wronskian conservation is consistent with QT [\ref{bib:vigierconf}].  This conservation arises from the Schwarzian derivative, which contains the quantum effects that transforms the CSHJE into the QSHJE.   On the other hand,  QEP conserves Bornian probability.  QEP finds that Schwarzian derivative term for and only for bound states implies the existence of an $L^2 (\mathbb{R})$ solution of the SSE [\ref{bib:fm}]. These $L^2 (\mathbb{R})$ solutions are compatible with the interpretation that the wave function has a Bornian probability amplitude.
As the denominator in Eq.\ (\ref{eq:gcm}) for $\partial W/\partial q$ is positive non-zero for finite $q$, $W$ is never a constant but a monotonic function of finite $q$ consistent with QEP [\ref{bib:fm},\ref{bib:fm3}]. For bound states, $\lim_{q \to \infty} \partial W/\partial q \longrightarrow 0$ asymptotically as exhibited numerically on Figures 1 and 3.   Such is also so for $\partial \widetilde{W}/\partial q$, cf. Figure 2. The behavior of asymptotically nullifying the conjugate momentum as $q \to \infty$ facilitates a finite change in a never-constant $W$ while transversing an infinite distance in $q$ [\ref{bib:fpl13}]. This asymptotic behavior manifests an asymptotic nodal singularity at $q=\pm \infty$.  For completeness, the conjugate momentum is not the mechanical momentum, $\partial_q W \ne m\dot{q}$ [\ref{bib:prd34},\ref{bib:fm},\ref{bib:fm3}].

The quantum reduced action is a generator of nonlocal motion.  Applying Jacobi's theorem, Eq.\ (\ref{eq:jacobitheorem}), generates the nonlocal quantum trajectory [\ref{bib:prd34},\ref{bib:rc},\ref{bib:fm}].  The equation of quantum motion for the quantum trajectory also parametrizes time while specifying the epoch, $\tau$, which is the temporal constant coordinate conjugate to energy.  Time parametrization by applying Jacobi's theorem to the general solution for the quantum reduced action, Eq.\ (\ref{eq:gw}), is given by

\begin{equation}
t-\tau = \frac{\partial W}{\partial E} = \frac{\hbar (\overbrace{AD-BC}^1)\, [\overbrace{(\partial_E \phi) \vartheta - (\partial_E \vartheta)
\phi}^{\mathcal{W}_E(\phi,\vartheta)}]}{(A^2+C^2)\phi^2 + 2(AB+CD)\phi \vartheta +(B^2+D^2)\vartheta^2}.
\label{eq:genjacobitheorem}
\end{equation}

\noindent  Microstate dependence of time parametrization is exhibited in Eq.\ (\ref{eq:genjacobitheorem}) by the coefficients $\{A,B,C,D\}$. The function $\mathcal{W}_E(\phi,\vartheta)$ in Eq.\ (\ref{eq:genjacobitheorem}) is analogous to the Wronskian where differentiation is with respect to $E$ instead of $q$.  For bound states in the $t,q$-plane $\lim_{q \to\infty} t - \tau \longrightarrow t_{\mbox{\scriptsize vertex}} - \tau$ asymptotically as exemplified numerically by Figures 4 and 5. The quantum trajectory for bound states at the infinite vertex in the open $t,q$-plane that forms an asymptotic cusp at $q \to \infty$ with asymptote $t=t_{\mbox{\scriptsize vertex}}$ [\ref{bib:prd26},\ref{bib:fpl13},\ref{bib:wyatt}].  A finite $t_{\mbox{\scriptsize vertex}}$ implies in this non-relativistic representation of quantum mechanics that the quantum trajectory transverses an infinite distance in a finite time [\ref{bib:prd26},\ref{bib:fpl13}].  The velocity of the quantum trajectory, $(\partial_q  \partial_E W)^{-1}$, becomes infinite at $q=\infty$ manifesting nonlocality.  In higher dimensions by the behaviors of the conjugate momentum, Eq.\ (\ref{eq:gcm}), and velocity of the quantum trajectory, Eq.\ (\ref{eq:jacobitheorem}), induce the quantum trajectorys' direction at the $q$-turning point, $q=\infty$, to be embedded in a surface of constant W because the wave normal becomes orthogonal to the trajectory as $q \to \infty$ [\ref{bib:fpl13}].  Again, Faraggi and Matone do not accept the validity of time parametrization by Eq.\ (\ref{eq:genjacobitheorem}) because of the ramifications of spatial compactification [\ref{bib:fm3}].  As Jacobi's theorem expressed in closed form, Eq.\ (\ref{eq:genjacobitheorem}), is well posed, time parametrization has L:ipschitz continuity.

Let us consider the QSHJE for a generic square well potential, $V$.  As the square well potential is piecewise continuous, each piecewise-continuous domain of the potential has its own solution, $W$. Functions and coefficients within the piecewise continuous, classically allowed region interior to the square well are denoted by the subscript 1; likewise within the classically forbidden region exterior to the square well, by the subscript 2.  The coefficients $\{A_1,B_1,C_1,D_1\}$ for the interior piecewise continuous segment of the square well are determined herein for a given energy, $E$, by coefficient normalization, Eq.\ (\ref{eq:coefnorm}), and by specifying the initial values $\{W,\partial_q W,{\partial^2}_q W\}|_{q=q_0}$ for the third-order QSHJE.  The initial point, $q_0$, for the square well is chosen to be the midpoint of its classically allowed region to take advantage of mirror symmetry inherent to the square well.  Another set of coefficients $\{A_2,B_2,C_2,D_2\}$ is needed in the exterior piecewise continuous segment in the classically forbidden region of the square well.   The set of coefficients $\{A_2,B_2,C_2,D_2\}$ is chosen consistent with $\{A_1,B_1,C_1,D_1\}$ so that the quantum reduced action maintains ${\mathcal{C}}^2$ continuity across the discontinuous potential steps of the square well.  For completeness and for a given energy $E$, the initial values for $\{W,\partial_q W,{\partial^2}_q W\}|_{q=q_0}$, and the coefficient normalization given by Eq. (\ref{eq:coefnorm}) are sufficient to specify a particular quantum reduced action (particular microstate).
\
\subsection{Quantization}

Let us now substantiate Jacobi's theorem, Eq.\ (\ref{eq:genjacobitheorem}), for time parametrization by investigating a finite potential well.  We consider a symmetric finite square-well potential, $V(q)$, given in natural units by

\begin{equation}
V(q) = \left\{ \begin{array}{cc}
               0 \   & \mbox{for } |q| \le a \\
               V_0 \ & \mbox{for } |q| > a
               \end{array}
               \right.
\label{eq:swv}
\end{equation}

\noindent where $V_0$ and $a$ are positive finite.  For quantized action variables, the energy is restricted in this investigation to $E \le V_0$.

The set of independent piecewise-continuous solutions $\{\phi,\vartheta\}$ of the SSE for the finite square-well potential given by Eq.\ (\ref{eq:swv}) contains real solutions.  The set $\{\phi_1,\vartheta_1\}$ is given by

\begin{equation}
\{ \phi_1,\vartheta_1 \} = \{ \sin(kq),\cos(kq) \} \ \ \mbox{for} \ \ -a \le q \le +a
\label{eq:pt1}
\end{equation}

\noindent where $k=(2mE)^{1/2}/\hbar.$  In the classically forbidden region denoted by the subscript 2, the set $\{\phi_2,\vartheta_2\}$ is given by

\begin{eqnarray}
\{\phi_2,\vartheta_2\} & = & \{ \sinh[\kappa(q-a)],\cosh[\kappa(q-a)] \} \  \mbox{if\ } q>+a  \nonumber \\
                    & = & \{ \sinh[\kappa(q+a)],\cosh[\kappa(q+a)] \} \  \mbox{if\ } q<-a
\label{eq:pt2}
\end{eqnarray}

\noindent where $\kappa = [2m(V_0 - E)]^{1/2}/\hbar$ and

\begin{equation}
k^2 + \kappa^2 = 2mV_0/\hbar^2 \ \ \mbox{for} \ \ E < V_0.
\label{eq:kappak}
\end{equation}

\noindent The chosen set $\{\phi_2,\vartheta_2\}$ is computationally convenient, for one has at the potential step
of the square well at $q=a$ that

\begin{equation}
\{\phi_2,\vartheta_2\}\big|_{q=a} = \{0,1\}\ \ \ \mbox{and}\ \ \ \{\partial_q \phi_2,\partial_q \vartheta_2\}\big|_{q=a} = \{\kappa,0\},
\label{eq:pt12i}
\end{equation}

\noindent which simplifies calculating the wave function that maintains $\mathcal{C}^1$ continuity for the SSE is second order.

First, we present the generator of the quantum motion, $W$, for a particular microstate of a bound state with energy $E$ as an heuristic example.  One constant of the quantum motion is the bound state energy, $E$.  The other constants that specify the particular microstate are the set of initial values at $q=0$, which specify a unique solution, $W$, of the QSHJE with the given energy, E [\ref{bib:prd29},\ref{bib:fpl9}].  We have chosen a set of initial values that renders symmetry to simplify the mathematics and allows us to investigate only a quarter-cycle of the orbital trajectory in phase space to determine the action variable, $J=4(W|_{q=\infty}-W|_{q=0})$.  The chosen set of initial values at the origin $q=0$ that solves the third-order QSHJE for the ground state is arbitrary
[\ref{bib:prd29}--\ref{bib:fm}].  Our choice here is

\begin{equation}
\{W_1,\partial_q W_1,{\partial^2}_q W_1\}|_{q=0}\  = \{0,\hbar k,0\}|_{q=0},
\label{eq:gsiv}
\end{equation}

\noindent which mimics the classical reduced action for the square well within the well itself (the classically allowed region).  This particular quantum reduced action, $W_1(E,0,\hbar k,0;q)$, which within the square well matches the initial values and Eq.\ (\ref{eq:coefnorm}), is antisymmetric [the symmetry of a bound-state $\psi$ and its microstate manifested by $W$ do not have to agree, cf. Eq.\ (\ref{eq:wasm}) and Figure 3]. This $W_1(0,\hbar k,0;E,q)$ is given by

\begin{equation}
W_1(0,\hbar k,0;E,q) = \hbar \arctan \left( \frac{A_1\phi_1(q)}{D_1\vartheta_1(q)} \right) = \hbar \arctan \left( \frac {\sin(kq)}{\cos(kq)}\right)
= {\hbar kq}\ \mbox{\ for\ } |q| \le a
\label{eq:gsw1}
\end{equation}

\noindent as expected.  Choosing $W_1|_{q=0} = 0$ and ${\partial^2}_q W|_{q=0} = 0$ as initial values ensures that $W$ be antisymmetric.

Inside the square well, the set of coefficients is given by $\{A_1,B_1,C_1,D_1\}=\{1,0,0,1\}$; the energy, $E=\hbar^2 k^2/(2m)$;
and the set of independent solutions, $\{\phi_1,\vartheta_1\}=\{\sin(kq),\cos(kq)\}$, are consistent with Eq.\ (\ref{eq:pt1}). The conjugate momentum inside the square well is given in natural units from Eq.\ (\ref{eq:gsw1}) by

\begin{equation}
\frac{\partial W_1}{\partial q} =  \hbar k = k \ \ \mbox{for } |q| \le a.
\label{eq:wq1}
\end{equation}

Let us investigate the behavior of the quantum reduced action, $W_2(0,\hbar k,0;E,q)$, in the classically forbidden region, $|q| > a$.  We now study the behavior of a general quantum reduced action with energy $E$ (not necessarily an energy eigenvalue) in the infinite limit, $q \to \infty$, of the classical forbidden region.  In the infinite limit of the quantum trajectory formulation, the ratios $\lim_{q \to +\infty} [\phi_2(q-a)/\vartheta_2(q-a)] = \lim_{q \to +\infty} \tanh[\kappa(q-a)]$ and  $\lim_{q \to +\infty}[\vartheta_2(q-a)/\phi_2(q-a)] = \lim_{q \to +\infty}\coth[\kappa(q-a)]$ have a common asymptote, 1, although they approach this asymptote from opposite directions.  Hence, the quantum reduced action in the infinite limit for the square well may be expressed using Eqs.\ (\ref{eq:gw}), (\ref{eq:fmw}), and (\ref{eq:pt2}) and with initial values given by Eq.\ (\ref{eq:gsiv}) as

\begin{equation}
\lim_{q \to \infty} W_2(\{W_1,\partial_q W_1,{\partial^2}_q W_1\}|_{q=0};E,q) = \lim_{q \to \infty} W_2(0,\hbar k,0;E,q)
      =  \hbar \arctan \left({\displaystyle \frac{A_2 + B_2}{C_2 + D_2}} \right).
\label{eq:gw2}
\end{equation}

\noindent If neither $A_2+B_2=0$ nor $C_2+D_2=0$, then the quantum reduced action in the infinite limit would be virtual, $\lim_{q \to \infty} (\widetilde{W})$. If either $A_2+B_2=0$ or $C_2+D_2=0$, then by symmetry

\[
J=4 \times \left[ \left( \lim_{q \to \infty} W_2(0,\hbar k,0;E,q) \right) - W_1(0,k,0;E,0) \right] = 4 \times \left( \lim_{q \to \infty} W_2(0,\hbar k,0;E,q) \right)
\]

\noindent where the chosen initial value for $W_1|_{q=0}=0$, cf.\ Eq.\ (\ref{eq:gsw1}). The coefficients $\{A_2,B_2,C_2,D_2\}$ are determined by the boundary values for $\mathcal{C}^2$ continuity at the potential step at $q=a$ and Eq. (\ref{eq:coefnorm}). Equation (\ref{eq:coefnorm}) ensures that both $A_2+B_2=0$ and $C_2+D_2=0$ cannot concurrently be true, otherwise the coefficients would be redundant.  Should $C_2+D_2=0$ while $A+B$ be $\pm$ finite, then, first,  the denominator of $\mathfrak{w}_2(+\infty)|_{C_2=-D_2}$, by Eq.\ (\ref{eq:frakw}) would be nil.  And second, $\mathfrak{w}_2(+\infty)|_{C_2=-D_2)}= +\infty$.  Consequently, $\lim_{q\to\infty}W_2$ by Eq.\ (\ref{eq:gw2}) would equal to $(2n-1)\hbar \pi/2,\ n=1,2,3,\cdots$ which is the action variable quantization for a quarter-cycle ($J/4$) of its symmetric orbit.  Also $W$ would be some microstate of the symmetric eigenfunctions, $\psi_{\mbox{\scriptsize symm}}$ of corresponding $n$.  Likewise, should $A_2+B_2=0$, then $\lim_{q\to\infty}W_2$ by Eq.\ (\ref{eq:gw2}) would equal $n\hbar\pi,\ n=1,2,3,\cdots$ which is the action variable quantization for a quarter-cycle ($J/4$) and $W$ would be would be some microstate of the antisymmetric eigenfunctions $\psi_{\mbox{\scriptsize anti}}$ of corresponding $n$.  These bound-state wave functions, $\psi_{\mbox{\scriptsize symm}}$ and $\psi_{\mbox{\scriptsize anti}}$ are $L^2(\mathbb{R})$ solutions.  As neither $A_2+B_2=0$ nor $C_2+D_2=0$ for a non-eigenvalue $\widetilde{E}$, then an $L^2(\mathbb{R})$ solution for the wave function would not exist for that $\widetilde{E}$.  Still, the solution for $\widetilde{\psi}$ of the SSE would exist and would be virtual.  Again, quantization of energy by QT is not based upon Bornian probability but upon Milne quantization.  The evaluation of $\lim_{q \to \infty} {W}_2$ is still given by Eq.\ (\ref{eq:gw2}) and is concurrently applicable to either bound-state or virtual conditions to render respectively either $J/4$ or  $\widetilde{J}/4$ for symmetric potentials.

We now examine the QEP representation of the quantum reduced action for bound states in the classically forbidden region.  Faraggi and Matone [\ref{bib:fm}] have found that the ratio $w=\psi^D/\psi$ in the extended $\hat{q}$-line, $-\infty \le q \le +\infty$ (again, QEP assumes compactification), is characterized by

\begin{equation}
w \ne \mbox{\itshape const},\ \ w \in  C^2,\ \ \mbox{and} \ \ {\partial^2}_q w \ \mbox{differentiable in } \mathbb{R},
\label{eq:qep1}
\end{equation}

\noindent and for and only for bound states

\begin{equation}
w(+\infty)= + \infty \mbox{\ \ \ and\ \ \ } w(-\infty)=-\infty.
\label{eq:qep2}
\end{equation}

\noindent Equations (\ref{eq:qep1}) and (\ref{eq:qep2}) are consistent with the condition for $w$ given by Faraggi and Matone [\ref{bib:fm3}] if and only if the set of independent solutions $\{\psi,\psi^D\}$ represents the wave solutions for a bound state where $\psi$ is $L^2(\mathbb{R})$ convergent and $\psi^D$ is divergent. As $w$ is locally invertible by $q \to q^{-1}$, then $W$ is a local homeomorphism and $\mathbb{R}$ is extended to $\hat{\mathbb{R}} = -\infty \cup \mathbb{R} \cup +\infty$ [\ref{bib:fm}]. Under a compactification assumption, QEP uses gluing $w(-\infty)$ to $w(+\infty)$ [\ref{bib:fm},\ref{bib:fpc}].

Thus for a bound state of the finite square well, the QEP analogy to Eq.\ (\ref{eq:gw2}) is given by Eq.\ (\ref{eq:fmw}) as

\begin{equation}
\lim_{q \to \pm \infty} W_2[(W_1,\partial_q W_1,{\partial^2}_q W_1)|_{q=0};E,q] =  \hbar \arctan \left({\displaystyle \frac{\mathfrak{A}_2 \times (\pm \infty) + \mathfrak{B}_2}{\mathfrak{C}_2 \times (\pm \infty) + \mathfrak{D}_2}} \right) \to \hbar \arctan \left({\displaystyle \frac{\mathfrak{A}_2}{\mathfrak{C}_2}}\right).
\label{eq:gw2fm}
\end{equation}

\noindent Should $\mathfrak{C}_2=0$, then $\lim_{q\to\infty}W_2 = (2n-1)\hbar \pi/2,\ n=1,2,3,\cdots$.  On the other hand, should $\mathfrak{A}_2=0$, then Eq.\ (\ref{eq:gw2fm}) would render $\lim_{q\to\infty}W_2 = n\hbar \pi,\ n=1,2,3,\cdots$.  Hence, QEP, by Eq.\ (\ref{eq:gw2fm}), and QT, by Eq.\ (\ref{eq:gw2}), render a mutually consistent action variable quantization.

Nevertheless, the quantum reduced action for non-eigenvalues is still well-behaved as shown by Eqs.\ (\ref{eq:gw2}) and (\ref{eq:gw2fm}).  As shown in Ref.\ \ref{bib:fpl9}, a quantization of action variable and its associated energy does not specify a particular set of initial values for the quantum reduced action: different initial values specify different microstates of the same bound $\psi$.  Again, Faraggi and Matone refer to these microstates as M\"{o}bius states [\ref{bib:fm}].

A comment on quantized asymmetric quantum action variables follows.  Let us return to Eq. (\ref{eq:wasm}) that quantizes in natural units the ground state action variable $J_{\mbox{\scriptsize gs}} = 2 \pi$ of the linear harmonic oscillator for an asymmetric $W_{\mbox{\scriptsize gs}}(0,1,0.5;0.5,q)$ in natural units.  However, in Eq.\ (\ref{eq:wasm}) at $q=-\infty$, $W_{\mbox{\scriptsize gs}}(0,1,0.5;0.5,-\infty) \approx W_{\mbox{\scriptsize gs}}(0.1,0.5;0.5,-10)  = -1.325\,817\,663\,668\,034$ instead of  equalling $-\pi/2$ for Eqs.\ (\ref{eq:gw2}) and (\ref{eq:gw2fm}) in natural units.  Likewise at $q=+\infty$ by Eq.\ (\ref{eq:wasm}), $W_{\mbox{\scriptsize gs}}(0,1,0.5;0.5,+\infty) \approx W_{\mbox{\scriptsize gs}}(0,1,0.5;0.5,+10) = +1.815\,774\,989\,921\,758$ instead of equalling $+\pi/2$ for Eqs.\ (\ref{eq:gw2}) and (\ref{eq:gw2fm}) in natural units.  This anomaly for asymmetric $W$s may be resolved by choosing in natural units the initial value $W_{\mbox{\scriptsize gs}}|_{q=0} = 0.2449\cdots = 0.244\,978\,663\,126\,86 = 1.815\,774\,989\,921\,758 - \pi/2 = \pi/2 - 1.325\,817\,663\,668\,034$.  In natural units with $W_{\mbox{\scriptsize gs}}|_{q=0} = 0.2449\cdots$, then $W(0.2449\cdots,1,0.5;0.5,\pm \infty)= \pm \pi/2$ in agreement with the behaviors of Eq.\ (\ref{eq:gw2}) of QT and Eq.\ (\ref{eq:gw2fm}) of QEP. The magnitude, $0.4899\cdots$, of twice that of $W_{\mbox{\scriptsize gs}}|_{q=0}$ is equal to the area between solid line and dashed line displayed on Fig.\ 3.

By symmetry, the quantization of the action variable may be determined by considering the behavior of the reduced action over a quarter-cycle along the positive $q$-axis.  Across the potential step of square well at $q=a$, the boundary values $\{W,\partial_q W,{\partial^2}_q W\}|_{q=a} = \{\hbar ka,\hbar k,0\}|_{q=a}$ must be matched by $W_1$ and $W_2$, for the QSHJE is a third-order partial differential equation.  The fourth condition for specifying the set of coefficients $\{A_2,B_2,C_2,D_2\}$, in the classically forbidden region $q>a$, is given by $A_2D_2-B_2C_2=1$, Eq.\ (\ref{eq:coefnorm}).  For the set of independent solutions $\{\sinh[\kappa (q-a)],\cosh[\kappa (q-a)]\}$ in the classically forbidden region $q> a$, the consequent values for the coefficients are given by Eqs.\ (\ref{eq:gw}), (\ref{eq:coefnorm}), and (\ref{eq:pt2}) as

\begin{equation}
A_2 = \left( \frac{k}{\kappa}\right)^{1/2} \cos(ka),\ \ B_2 = \left( \frac{\kappa}{k}\right)^{1/2} \sin(ka),\ \ C_2 = -\left( \frac{k}{\kappa}\right)^{1/2} \sin(ka),\ \ \mbox{and}\ \ D_2 = \left( \frac{\kappa}{k}\right)^{1/2} \cos(ka).
\label{eq:abcd2}
\end{equation}

\noindent  The values of the coefficients maintain $\mathcal{C}^2$ continuity for the quantum reduced action across the potential step of the
square well at $q=a$.  These coefficients are composed of relationships among $k,\ \kappa,$ and $a$ that keep the coefficients dimensionless.  The coefficients remain piecewise constants in the classically forbidden region.

The set of coefficients $\{A_2,B_2,C_2,D_2\}$ specifies the microstate in the classically forbidden region [\ref{bib:fpl9}], and also specifies the terminal value of the quantum reduced action at $+\infty$  by Eq.\ (\ref{eq:gw2}).  The quantum reduced action for an eigenvalue energy displays critical point behavior of the asymptotic stable nodal singularity class as $q \to +\infty$ regardless of microstate specification [\ref{bib:prd25},\ref{bib:bo}].  The analogy of this behavior in QEP is the requirement that $W \ne const$ [\ref{bib:fm}].

We now confirm that the set of coefficients $\{A_2,B_2,C_2,D_2\}$  maintains $\mathcal{C}^2$ continuity for $W$ across the potential step at $q=a$.   In the classically forbidden region $q>a$, the quantum reduced action $W_2$ is given for the square well by Eqs.\ (\ref{eq:gw}), (\ref{eq:pt2}) and
(\ref{eq:abcd2}) as

\begin{equation}
W_2(0,\hbar k,0;E,q) = \hbar \arctan \left( \frac{\left(\frac{k}{\kappa}\right)^{1/2} \cos(ka)\,\sinh[\kappa(q-a)]\, +\,  \left( \frac{\kappa}{k}\right)^{1/2} \sin(ka)\,\cosh[\kappa(q-a)]}{-\left( \frac{k}{\kappa}\right)^{1/2} \sin(ka)\,\sinh[\kappa(q-a)]\, +\,  \left( \frac{\kappa}{k}\right)^{1/2} \cos(ka)\,\cosh[\kappa(q-a)]}\right)\ \ \mbox{for}\ q>a.
\label{eq:gsw2}
\end{equation}

\noindent At the potential step of the square well, $q=a$, the quantum reduced actions, $W_1$ and $W_2$, match each other.  One has by Eqs.\ (\ref{eq:pt12i}),
(\ref{eq:gsw1}) and (\ref{eq:gsw2}) that

\begin{equation}
W_2(0,k,0;E,a) = \hbar \arctan(B_2/D_2) = \hbar ka = W_1(0,k,0;E,a).
\label{eq:wi12}
\end{equation}

\noindent The continuity of the quantum reduced action, Eq.\ (\ref{eq:wi12}), implies the continuity of $\mathfrak{w}$ across the potential step.

The conjugate momentum in the classically forbidden region is given by Eq.\ (\ref{eq:gsw2}) as

\begin{equation}
\frac{\partial W_2(0,\hbar k,0;E,q)}{\partial q} = \frac{ \hbar \kappa}{\frac{k}{\kappa}\sinh^2[\kappa(q-a)] + \frac{\kappa}{k}\cosh^2[\kappa(q-a)]} \ \ \mbox{for}\ q>a.
\label{eq:gcm2}
\end{equation}

\noindent For all finite $q>a$, the conjugate momentum remains finite, which is consistent with Faraggi and Matone's QEP requirement that the quantum reduced
action may not be constant [\ref{bib:fm},\ref{bib:fm3}].  Again, the QT analogy is that the QSHJE displays critical point behavior of the asymptotic stable nodal singularity
class as $q \to +\infty$, where the conjugate momentum goes to zero regardless of microstate specification [\ref{bib:prd25},\ref{bib:bo}].

At the potential step of the square well, the conjugate momenta, $\partial_q W_1$ and $\partial_q W_2$, match each other.  From Eqs.\ (\ref{eq:pt12i}),
(\ref{eq:wq1}), and (\ref{eq:gcm2}) one has at $q=a$ that

\begin{equation}
\frac{\partial W_2}{\partial q}\bigg|_{q=a} = \hbar k = \frac{\partial W_1}{\partial q}\bigg|_{q=a}.
\label{eq:cmi12}
\end{equation}

We now examine the continuity of ${\partial^2}_qW$ across the step in the square well potential at $q=a$.  In the classically allowed region, $-a < q < a$,
the reduced action, $W_1$, is linear in $q$, Eq.\ (\ref{eq:gsw2}).  Hence, ${\partial^2}_q W_1|_{q=a} = 0$.  In the classically forbidden region, $q \ge a$,
the conjugate momentum, $\partial_q W_2$, has a one-sided maximum at $q=a$.  Further analysis shows that in the conjugate momentum's denominator both $\cosh^2[\kappa(q-a)]$
and $\sinh^2[\kappa(q-a)]$ form at $q=a$ non-negative, one-sided minima and their coefficients $k/\kappa$ and $\kappa/k$ are positive real. Hence, the one-sided
${\partial^2 W_2}|_{q=a} = 0$ also.  Thus, ${\partial^2}_q W_1$ and ${\partial^2}_q W_2$ match at $q=a$.  Note that the continuity of
$\{W,\partial_q W,{\partial^2}_q W\}|_{q=a}$ is applicable to non-eigenvalues of energies and action variables.

Let us now investigate bound states.  For the ground state and all other symmetric, excited bound states whose $W_2$ is described by Eq.\ (\ref{eq:gsw2}),\ then we have by Eq.\ (\ref{eq:abcd2}) that
\begin{equation}
-\frac{C_2}{D_2}\ \ =\ \ \underbrace{\frac{k}{\kappa} \tan(ka)\ \ = \ \ 1}_{\stackrel{\mbox{\scriptsize square well}}{\mbox{\scriptsize $\psi_{\mbox{\tiny symmetric}}$ quantization}}}.
\label{eq:qsymm}
\end{equation}

\noindent where

\begin{equation}
\sin(ka) = \pm \frac{\kappa}{(k^2 + \kappa^2)^{1/2}} \ \ \ \mbox{and} \ \ \ \cos(ka) = \pm \frac{k}{(k^2 + \kappa^2)^{1/2}}
\label{eq:trigsymm}
\end{equation}

\noindent where in turn the $\cos(ka)$ and $\sin(ka)$ have the same sign.  Thus, the principal value of the argument $ka$ must be in either the first or third quadrant.  As

\[
-C_2 = D_2 = \left( \frac{k \kappa}{(k^2 + \kappa^2)} \right)^{1/2} =  \hbar \left( \frac{k \kappa}{2mV_0} \right)^{1/2}
\]

\noindent in Eq.\ (\ref{eq:qsymm}), the quantizing condition for symmetric bound states for the square well may be re-expressed by

\begin{equation}
C_2 + D_2 = 0
\label{eq:qsymmalt}
\end{equation}

\noindent consistent with the symmetric quantizing conditions for Eq.\ (\ref{eq:gw2}).  Then the denominator, $\Theta_{2,\psi_{\mbox{\tiny symm}}}$, of the bilinear transformation $\mathfrak{w}$ of Eq.\ (\ref{eq:frakw}) or alternately of the argument of the arc tangent function of $W_{2,\psi_{\mbox{\tiny symm}}}$ (always antisymmetric even for $\psi_{\mbox{\scriptsize symmetric}}$) of Eq.\ (\ref{eq:gsw2}) becomes

\begin{equation}
\Theta_{2,\psi_{\mbox{\tiny symm}}} = -\hbar \left( \frac{k \kappa}{2mV_0} \right)^{1/2} \sinh[\kappa (q-a)] + \hbar \left( \frac{k \kappa}{2mV_0} \right)^{1/2} \cosh[\kappa (q-a)] = \hbar \left( \frac{k \kappa}{2mV_0} \right)^{1/2} \exp[-\kappa (q-a)].
\label{eq:hyposymm}
\end{equation}

\noindent  Equation (\ref{eq:hyposymm}) is a familiar hyperbolic identity with the scaling factor $\hbar [k \kappa/(2mV_0)]^{1/2}$.  Equation (\ref{eq:hyposymm}) by the superpositional principle also converts $\Theta_{2,\psi_{\mbox{\tiny symm}}}$ from a hyperbolic representation, $\{\phi_2,\vartheta_2\}$, to an exponential representation, $\{\phi_2+\vartheta_2,\phi_2-\vartheta_2\} = \{\exp[+\kappa(q-a)],\exp[-\kappa(q-a)]\},\ q>a$,  in which the quantizing condition eliminates the positive exponential component, $\exp[\kappa(q-a)]$.

Likewise, the values of the coefficients $A_2$ and $B_2$ for symmetric bound states are given by
\[
A_2 = \hbar \left( \frac{k^3}{\kappa\ 2mV_0} \right)^{1/2} \ \ \mbox{and} \ \ B_2 = \hbar \left( \frac{\kappa^3}{k\ 2mV_0} \right)^{1/2}
\]

\noindent where $A_2D_2 - B_2C_2 = 1$.  To show that the symmetric bound state $J_{\psi_{\mbox{\tiny symm}}}$ is consistent with Faraggi and Matone's $L^2(\mathbb{R})$ requirement, we express $W_{2,\psi_{\mbox{\tiny symm}}}$
 by a set of independent exponential solutions $\{\exp[+\kappa(q-a)],\exp[-\kappa(q-a)]\},\ q>a$.  The ground state and other symmetric bound states with energy $E_{\psi_{\mbox{\tiny symm}}}$ by Eqs.\ (\ref{eq:abcd2}), (\ref{eq:qsymm}), (\ref{eq:qsymmalt}), and (\ref{eq:hyposymm}) may have their quantum reduced actions expressed as

\begin{eqnarray}
W_{2,\psi_{\mbox{\tiny symm}}}(0,\hbar k,0;E_{\psi_{\mbox{\tiny symm}}},q)& = &\hbar \arctan\left( \frac{ \frac{A_2 - B_2}{2} \exp[-\kappa (q-a)] + \frac{A_2 + B_2}{2} \exp[\kappa(q-a)]}{\frac{C_2 - D_2}{2} \exp[-\kappa(q-a)] + {\frac{ C_2 + D_2}{2} \exp[\kappa (q-a)]}}\right) \label{eq:w2symm1} \\
                                        & = &\hbar \arctan \left[ \frac{k}{\kappa} \left( \frac{\kappa^2 + k^2}{2 \kappa^2} \exp[2\kappa(q-a)] + \frac{\kappa^2-k^2}{2\kappa^2} \right) \right].  \label{eq:w2symm2}
\end{eqnarray}

\noindent  This implies that as $q \to +\infty$, the quantum reduced action for $E_{\psi_{\mbox{\tiny symm}}}$ is evaluated by Eq.\ (\ref{eq:w2symm2}) to be  $W_2(0,k,0;E_{\psi_{\mbox{\tiny symm}}},\infty) \to (n-1/2)\pi\hbar = (n-1/2)h/2,\ n=1,2,3,\cdots$  as expected.  Hence, $J_{\psi_{\mbox{\tiny symm}}}=(4n-2)\pi\hbar = (2n-1)h,\ n=1,2,3,\cdots$.  The behavior of the denominator, $\Theta_{2,\psi_{\mbox{\tiny symm}}}$, has already been investigated by Eq.\ (\ref{eq:hyposymm}). The function $\Theta_{2,\psi_{\mbox{\tiny symm}}}$ is also a solution of the associated SSE in the classically forbidden region $q>a$ by the superpositional principle for ordinary differential equations.  As such, $\Theta_{2,\psi_{\mbox{\tiny symm}}}(q)$ diminishes exponentially as $q \to +\infty$. This implies the existence of a bound-state solution of the SSE for symmetric states, which is also an $L^2(\mathbb{R})$ solution on the extended $\hat{q}$-line consistent with Faraggi and Matone's QEP for bound states.  QEP infers Bornian probability and quantization of $E$ from $\Theta_{2,\psi_{\mbox{\tiny symm}}}(q)$ being an $L^2(\mathbb{R})$ solution for symmetric bound states, while QT quantizes $E$ from Milne quantization, $J_{\psi_{\mbox{\tiny symm}}} = (2n-1)h,\ n=1,2,3,\cdots$.  Likewise for non-eigenvalue $\widetilde{E}$, then $C_2+D_2$ would be finite.  Hence, $\Theta_2$ would contain a term $(C+D)\exp[\kappa (q-a)]$ that would be finite throughout the classically forbidden region, $a<q \le \infty$.

Let us now examine the behavior of the quantum reduced action for antisymmetric bound states of the square well.  These states are those excited states with
antisymmetric wave functions,$\psi_{\mbox{\scriptsize anti}}$s, and have action variables given by $J_{\psi_{\mbox{\tiny anti}}} = 4n \pi\hbar = 2nh,\ n=1,2,3,\cdots$.  The examination of antisymmetric bound state is analogous to that for the symmetric bound state.  The coefficients $A_2$ and $B_2$ of $W_2$ that are consistent with antisymmetric bound states by Eq.\ (\ref{eq:abcd2}) have the relationship given by

\begin{equation}
\frac{A_2}{B_2}\ \ =\ \ \underbrace{+\frac{k}{\kappa} \cot(ka)\ \ =\ \ -1}_{\stackrel{\mbox{\scriptsize square well}}{\mbox{\scriptsize $\psi_{\mbox{\tiny antisymmetric}}$ quantization}}}
\label{eq:qanti}
\end{equation}

\noindent  where

\begin{equation}
\cos(ka) = \pm \frac{\kappa}{(k^2 + \kappa^2)^{1/2}} \ \ \ \mbox{and} \ \ \ \sin(ka) = \mp \frac{k}{(k^2 + \kappa^2)^{1/2}}
\label{eq:triganti}
\end{equation}

\noindent where in turn the $\cos(ka)$ and $\sin(ka)$ have opposite signs. [Note that the quantization formulas, Eqs.\ (\ref{eq:qsymm}) and (\ref{eq:qanti}), are derived without using the fact that the bound-state $\psi$ is $L^2(\mathbb{R})$.]  That is the principal value of the argument $ka$ must be in either the second or fourth quadrant.  As

\[
-A_2 = B_2 = \left( \frac{k \kappa}{(k^2 + \kappa^2)} \right)^{1/2} =  \hbar \left( \frac{k \kappa}{2mV_0} \right)^{1/2}
\]

\noindent in Eq.\ (\ref{eq:qanti}), the quantizing condition for antisymmetric bound states for the square well may be re-expressed by

\begin{equation}
A_2 + B_2 = 0
\label{eq:qasymmalt}
\end{equation}

\noindent consistent with the antisymmetric quantizing conditions of Eq.\ (\ref{eq:gw2}).  In a manner analogous to the treatment of $\Theta_{2,\psi_{\mbox{\tiny symm}}}$ in Eq.\ (\ref{eq:hyposymm}),  the numerator, $\Phi_{2,\psi_{\mbox{\tiny anti}}}$ of the argument, $\mathfrak{w}$, of the arc tangent function of $W_{2,\psi_{\mbox{\tiny anti}}}$ of Eq.\ (\ref{eq:gsw2}), becomes

\begin{equation}
\Phi_{2,\psi_{\mbox{\tiny anti}}} = -\hbar \left( \frac{k \kappa}{2mV_0} \right)^{1/2} \sinh[\kappa (q-a)] + \hbar \left( \frac{k \kappa}{2mV_0} \right)^{1/2} \cosh[\kappa (q-a)] = \hbar \left( \frac{k \kappa}{2mV_0} \right)^{1/2} \exp[-\kappa (q-a)],
\label{eq:hypoanti}
\end{equation}

\noindent which is another hyperbolic identity.  Equation (\ref{eq:hypoanti}) by superposition changes the hyperbolic solutions of the SSE in the classically forbidden region of the square well into an exponential solution, $\exp[-\kappa(q-a)],\ q>a$.

Likewise, the values of the coefficients $C_2$ and $D_2$ for antisymmetric bound states are given by
\[
C_2 = -\hbar \left( \frac{k^3}{\kappa\ 2mV_0} \right)^{1/2} \ \ \mbox{and} \ \ D_2 = -\hbar \left( \frac{\kappa^3}{k\ 2mV_0} \right)^{1/2}
\]

\noindent where $A_2D_2 - B_2C_2 = 1$.  To show that the antisymmetric bound state $J_{\psi_{\mbox{\tiny anti}}}$ is consistent with Faraggi and Matone's $L^2(\mathbb{R})$ requirement, we express $W_{2,\psi_{\mbox{\tiny anti}}}$ by a set of independent exponential solutions $\{\exp[+\kappa(q-a)],\exp[-\kappa(q-a)]\},\ q>a$.     Thus, the antisymmetric bound states with energy $E_{\psi_{\mbox{\tiny anti}}}$ have quantum reduced actions given by

\begin{equation}
W_{2,\psi_{\mbox{\tiny anti}}}(0,\hbar k,0;E_{\psi_{\mbox{\tiny anti}}},q) = \hbar \arctan \left[ \frac{k}{\kappa} \left( \frac{\kappa^2 + k^2}{2 \kappa^2} \exp[2\kappa(q-a)] +
                                                   \frac{\kappa^2-k^2}{2\kappa^2} \right)^{-1} \right],  \label{eq:w2anti2}
\end{equation}

\noindent The quantum reduced actions for excited antisymmetric wave functions, $\psi_{\mbox{\scriptsize anti}}$, are evaluated at $q \to \infty$ as $W_2 = n \hbar \pi,\ n=1,2,3,\cdots$.  Consequently $J_{\psi_{\mbox{\tiny anti}}} = 4n\hbar\pi = 2nh,\ n=1,2,3,\cdots$.  The argument, $\mathfrak{w}$, of the arc tangent function of $W_{2,\psi_{\mbox{\tiny anti}}}$ has vertical asymptotes manifesting a jump to the next Riemann sheet.   The numerator, $\Phi_{2,\psi_{\mbox{\tiny anti}}}$ of the argument, $\mathfrak{w}$, of the arc tangent of the first excited state is given by

\begin{equation}
\Phi_{2,\psi_{\mbox{\tiny anti}}}(q) = A_2 \phi_2(q) + B_2 \vartheta_2(q) = A_2 \phi_2(q) - A_2 \vartheta_2(q) = A_2 \exp[-\kappa (q-a)],\ \ \mbox{for}\ \ q>a,
\label{eq:Phi}
\end{equation}

\noindent which for $q>a$ is also the exponentially decaying solution of the associated SSE by the superpositional principle.  As such, $\Phi_{2,\psi_{\mbox{\tiny anti}}}(q)$, analogous to the ground state $\Theta_{2,\psi_{\mbox{\tiny symm}}}(q)$, also decays exponentially as $q \to +\infty$.  This implies that the antisymmetric solution, $\Phi_2(q)$, of the SSE for a first, third, fifth, $\dots$ excited state is an $L^2(\mathbb{R})$ solution on the extended $\hat{q}$-line consistent with Faraggi and Matone's QEP for bound states.

Forsyth [\ref{bib:arf}] and Hecht and Mayer [\ref{bib:hm}] did not require that the quantum reduced actions be eigenfunctions of QSHJEs to map a quantum reduced action from one QSHJE into the quantum reduced action of a different QSHJE.  The quantum trajectory representation does not disqualify a $\widetilde{\psi}$ from mapping into other $\psi$s or $\widetilde\psi$s by a Schwarzian derivative process.

We may now express Eq.\ (\ref{eq:gw2}) for general quantum reduced action for $q \to \infty$ where the coefficients $\{A_2,B_2,C_2,D_2\}$ are expressed in terms
of $k$ acting as energy's proxy. The microstate $W(E,0,\hbar k,0;q)$ for eigenvalue energy $E$ of the quantum square well is used as an example.  The quantum reduced action of Eq. (\ref{eq:gw2}) may be re-expressed for this microstate by Eq.\ (\ref{eq:abcd2}) as [\ref{bib:dwight}]

\begin{equation}
\lim_{q \to \infty} W_2(0,\hbar k,0;E,q) =  \hbar \arctan \left({\displaystyle \frac{A_2 + B_2}{C_2 + D_2}} \right)
      =  \hbar \arctan \bigg(\frac{\sin[\overbrace{ka+\mbox{arccot}(\kappa/k)}^{\mbox{\tiny anti quantization if 0}}]}
       {\sin[\underbrace{ka-\arctan(\kappa/k)}_{\mbox{\tiny symm quantization if 0}}]}\bigg).
\label{eq:gwquant}
\end{equation}

\noindent Again, the quantization condition for the quantum square well arises naturally as it did in Eqs.\ (\ref{eq:qsymm}) and (\ref{eq:qanti}).  When energy is such that the symmetric quantization is fulfilled, then $\sin[ka-\arctan(\kappa/k)] = 0$ and $W_2|_{q=\infty}=J_{\psi_{\mbox{\tiny symm}}}/4=(2n-1)\pi\hbar/2=(2n-1)h/4,\ n=1,2,3,\cdots$.  If energy is such that the antisymmetric quantization is fulfilled, then $\sin[ka+\mbox{arccot}(\kappa/k)]=0$ and $W_2|_{q=\infty}=J_{\psi_{\mbox{\tiny anti}}}/4=n\pi\hbar=nh/2,\ n=1,2,3,\cdots$.  Bound state quantization is manifested for  QT by quantizing the  quantum action variable, $J=2n\pi\hbar=nh,\ n=1,2,3,\cdots$.  All other energies less than $V_0$ will be virtual as the argument, $\mathfrak{w}$, of the arc tangent function in Eq.\ (\ref{eq:gwquant}) would be non-zero and finite, which would render a virtual $\widetilde{J}$.

A finite square well has a finite, positive number of bound states.  QT and QEP can both find this number.  The upper bound for $k$ is $k_{\mbox{\scriptsize ub}} = (2mV_0)^{1/2}/\hbar$, for which $\kappa = 0$ by Eq.\ (\ref{eq:kappak}).  For initial values given by Eq.\ (\ref{eq:gw2}) and from Eq.\ (\ref{eq:abcd2}), the upper bound for the action variable $J_{\mbox{\scriptsize ub}}$ is given by

\begin{eqnarray}
\frac{J_{\mbox{\scriptsize ub}}}{4} = W(0,k_{\mbox{\scriptsize ub}},0;V_0,\infty) & = & \hbar \arctan\left(\frac{A_2}{C_2} \right) = \hbar \arctan\left[-\cot\left(\frac{(2mV_0)^{1/2}}{\hbar}a\right)\right] \nonumber \\
& = & \hbar \arctan\left[\tan\left(\frac{(2mV_0)^{1/2}}{\hbar}a + \frac{\pi}{2} \right) \right] = (2mV_0)^{1/2}a + \frac{h}{4}.
\label{eq:maxn}
\end{eqnarray}

\noindent  The number of allowed bound states for $J_{\mbox{\scriptsize ub}}$ is given by

\begin{equation}
\left[4 \frac{(2mV_0)^{1/2}a}{h} +1\right]
\label{eq:maxni}
\end{equation}

\noindent where $[\chi]$ denotes the largest integer equal to or less than $\chi$.  The number of bound states derived by Eq.\ (\ref{eq:maxni}) is                                                                                                                                                                                                                                                                                                                                                                                                                                                                                                                                                                                                                                                                                                                                                                                                                                                                                                                                                                                                                                                                                                                                                                                                                                                                                                                                                                                                                                                                                                                                                                                                                                                                                                                                                                                                                                                                                                                                                                                                                                                                                                                                                                                                                                                                                                                                                                                                                                                                                                                                                                                                                                                                                                                       consistent with those derived by the $\psi$-representation [\ref{bib:eisberg}].  All finite square wells, as all one-dimensional potential wells, have at least one bound state by Eqs.\ (\ref{eq:maxn}) and (\ref{eq:maxni}).  If in the limit $V_0 \to 0$ (no longer a finite square well), then by Eq.\ (\ref{eq:maxni}) the solitary remaining bound state will have $E=0$, $\kappa=0$, and QEP-defined $\mathfrak{W}(q) \equiv E-V(q)=0$ perforce for all $q$.  While under QEP, there exists a coordinate transformation  that maps any physical state to one where $\mathfrak{W}=0$ [\ref{bib:fm},\ref{bib:fm3}], again Forsyth [\ref{bib:arf}] and Hecht and Mayer [\ref{bib:hm}] have not required that a state be physical for mapping by a Schwarzian derivative process.

The development for quantizing the action variable did not need Bornian probability.  Bornian probability was only adduced in \S3.2 to support QEP's energy quantization.

For completeness, the quantization of the action variable for bound states of a general potential has been shown elsewhere by contour integration [\ref{bib:prd34}].

The QT results of this subsection could have been achieved by QEP.

\subsection{Continuity of Action Variable with respect to Energy}

Continuity of the quantum reduced action with regard to energy is numerically maintained for the linear harmonic oscillator as shown in Tables 1 and 2 of \S 2.1. We substantiate this by investigating the finite square well for the continuity of the quantum reduced action with energy by establishing an analytic relationship between action variable and energy's proxy $k$ that is valid for virtual $\widetilde{J}$ and $\widetilde{E}$ as well as eigenvalue $J$ and $E$.

Let us assume that at least two bound states exist for the finite quantum square well. As previously discussed in \S2.1, the domains of $\widetilde{E}$ and $\widetilde{J}$ are segmented by limit points, eigenvalue $E$s and $J$s.  We shall examine for the finite square well the continuity of $\widetilde{J}$ as $\widetilde{E}$ progresses from the ground state $E_{\mbox{\scriptsize gs}}$ to the first excited state $E_{\mbox{\scriptsize fes}}$ analytically.  The choice of the initial values for the quantum reduced action specifying the particular microstate is arbitrary.  Any $\widetilde{E}$ given by $E_{\mbox{\scriptsize gs}} < \widetilde{E} < E_{\mbox{\scriptsize fes}}$ will render a $\widetilde{J}$ that obeys $2\pi\hbar < \widetilde{J} < 4\pi\hbar$ as exemplified on Table 1. If this were not true, than some  $\widetilde{E}$ given by $E_{\mbox{\scriptsize gs}} < \widetilde{E} <E_{\mbox{\scriptsize fes}}$ would render a $J$ that equals either $2\pi\hbar$ or $4\pi\hbar$.  In either case, an eigenvalue $J$ would be associated with two distinct energies, an impossibility.

We now investigate in closed form the quantum reduced action at $q \to \infty$ as a proxy for the action variable where symmetry permits $J/4 = [(\lim_{q \to \infty} W_2) - W_1]|_{q=0}$ and where in turn $W_1|_{q=0}=0$.  For the finite square well with a set of initial values $\{0,\hbar k,0\}|_{q=0}$ given by Eq.\ (\ref{eq:gsiv}), the action variable and quantum reduced action in the limit $q \to \infty$ are given by Eqs.\ (\ref{eq:gsw1}), (\ref{eq:gw2}) and (\ref{eq:abcd2}) as

\begin{equation}
\frac{J}{4} = \left( \lim_{q \to \infty} W_2(0,\hbar k,0;E,q) \right) - \underbrace{W_1(0,\hbar k,0;E,0)}_0 = \hbar \arctan \left( \frac{ \left( \frac{k}{\kappa}\right)^{1/2} \cos(ka) + \left( \frac{\kappa}{k}\right)^{1/2} \sin(ka)}{-\left( \frac{k}{\kappa}\right)^{1/2} \sin(ka) + \left( \frac{\kappa}{k}\right)^{1/2} \cos(ka)} \right).
\label{eq:gswrai}
\end{equation}

\noindent Equation (\ref{eq:gswrai}) may be rewritten as

\begin{equation}
\left[ \cos\left(\frac{J}{4\hbar}\right) \kappa+\sin\left(\frac{J}{4\hbar}\right)k \right] \sin(ka) = \left[ -\cos\left(\frac{J}{4\hbar}\right) k + \sin\left(\frac{J}{4\hbar}\right) \kappa \right] \cos(ka).
\label{eq:univquant}
\end{equation}

\noindent  As $k^2+\kappa^2=2mV_0/\hbar^2$, Eq.\ (\ref{eq:kappak}) for $E \le V_0$, Eq.\ (\ref{eq:univquant}) may be expressed by just two variables: the action variable $J$, and the wave number $k$ acting as the proxy for $E$.  Thus for a given set of initial values $\{0,\hbar k,0\}|_{q=0}$, the transcendental Eq.\ (\ref{eq:univquant}) establishes a reversible mapping between continuous values of $J$ and $k$, $E$'s proxy.

For the ground state action variable where $J_{\mbox{\scriptsize gs}}=2\pi\hbar=h$, Eq.\ (\ref{eq:univquant}) reduces to $k\tan(ka)=\kappa$, which is the quantization condition for the ground state and other symmetric excited bound states of the square well, Eq.\ (\ref{eq:qsymm}). For the first excited state  action variable where $J_{\mbox{\scriptsize fes}}=4\pi\hbar=2h$, Eq.\ (\ref{eq:univquant}) reduces to $-k \cot{ka}=\kappa$, which is the quantization condition for antisymmetric excited states of the square well, Eq.\ (\ref{eq:qanti}).

Other action variables that are not equal to $n\pi\hbar,\ n=2,4,6,\cdots$, are virtual.
Any $\widetilde{J}$ induces a particular $\widetilde{E}$ by Eqs.\ (\ref{eq:gswrai}) and (\ref{eq:univquant}) that is dependent upon the initial values for the quantum reduced action.  The variations of the initial values for $\widetilde{W}$ induce a dispersion in the relationship between $\widetilde{J}$ and $\widetilde{E}$ as exhibited for the linear harmonic oscillator by Table 1.  On the other hand, variations in the initial values for $W$ do not induce any dispersion in the relationship between the eigenvalues of $J$ and $E$ as also exhibited by Table 1.  Nevertheless, the orbit of the quantum trajectory in phase space for a bound state eigenvalues of $E$ or $J$ is still dependent upon the initial values of $W$.  This dependency manifests the existence of microstates, which again QEP has identified as M\"{o}bius states [\ref{bib:prd26}--\ref{bib:rc}].

There is a simplification to Eq.\ (\ref{eq:univquant}) that facilitates resolving a relationship between $\widetilde{J}$ and $\widetilde{E}$ or its proxy $k$ for certain finite square wells with finite potential step $V_0$ under particular circumstances. These circumstances include using the initial values for $\widetilde{W}$ given by Eq.\ (\ref{eq:gsiv}), $\{0,\hbar k,0\}|_{q_0}$.  Let us consider the cases $\widetilde{J}_n = (2n-1)\pi\hbar = (2n-1)h/2,\ n=1,2,3,\cdots$.  Physically, these $\widetilde{J}$s are midway between their immediate neighboring eigenvalue $J$s except for $n=1$ for which $\widetilde{J}=J_{\mbox{\scriptsize gs}}/2$.  Let us examine for mathematical convenience only those cases where $k=\kappa=(mV_0)^{1/2}/\hbar$ by Eq.\ (\ref{eq:kappak}). Hence $(k/\kappa)^{\pm 1/2} = 1$.  It follows that $\widetilde{E}=V_0/2$ where $\widetilde{E}$ as well as $V_0$ are fixed independent of $n$.   The only remaining parameter in Eq.\ (\ref{eq:univquant}) is the finite half-width of the finite square well, $a_n$, which can be determined in terms of the other assumed parameters (this selection is arbitrary; $a$ could be fixed and $V_0$ could be determined in terms of the other parameters).  The principle values of $\widetilde{J}_n/(4\hbar)$ mod $2\pi$ form the set $\{\pi/4,3\pi/4,5\pi/4,7\pi/4\}$.  Then $|\sin[\widetilde{J}_n/(4\hbar)]|,|\cos[\widetilde{J}_n/(4\hbar)]|=2^{-1/2}$. If $\sin[\widetilde{J}_n/(4\hbar)]$ and $\cos[\widetilde{J}_n/(4\hbar)]$ have the same sign (first and third quadrants), then $n=1,3,5,\cdots$ and the right side of Eq.\ (\ref{eq:univquant}) is nullified.  This in turn reduces Eq.\ (\ref{eq:univquant}) to $\pm 2^{1/2}\sin{ka_n}=0$ or $ka_n=(n+1)\pi/2$. The allowable half width of the finite square well, $a_n$, may be expressed in terms of the other parameters by Eq.\ (\ref{eq:univquant}) as

\begin{equation}
a_n = \frac{n\pi}{2k} = \frac{n+1}{2}\ \frac{\pi\hbar}{(mV_0)^{1/2}} = \frac{n+1}{4n-2}\ \frac{\widetilde{J}_n}{(mV_0)^{1/2}},\ \ n=1,3,5,\cdots.
\label{eq:aeven}
\end{equation}

\noindent Note that the solution $a_n=0$ for $\sin(ka_n)=0$ is inconsistent with a finite square well. On the other hand, if $\sin[\widetilde{J}_n/(4\hbar)]$ and $\cos[\widetilde{J}_n/(4\hbar)]$ have opposite signs (second and fourth quadrants), then $n=2,4,6,\cdots$ and the left side of Eq.\ (\ref{eq:univquant}) is nullified.  This in turn reduces Eq.\ (\ref{eq:univquant}) to $\mp 2^{1/2}\cos{ka_n}=0$ or $ka_n=(n+1)\pi/2$. The allowable half width, $a_n$, may be expressed in terms of the other parameters as

\begin{equation}
a_n = \frac{n\pi}{2k} = \frac{n+1}{2}\ \frac{\pi\hbar}{(mV_0)^{1/2}} = \frac{n+1}{4n-2}\ \frac{\widetilde{J}_n}{(mV_0)^{1/2}},\ \ n=2,4,6,\cdots.
\label{eq:aodd}
\end{equation}

\noindent Equations (\ref{eq:aeven}) and (\ref{eq:aodd}) exhibit the same progressive increase in $a_n$ with increasing $n$ whether odd or even. Equations (\ref{eq:aeven}) and (\ref{eq:aodd}) can be consolidated where $n=1,2,3,\cdots$.  Each $a_n$ in combination with $V_0$ specifies a unique finite square well.  The transcendental Eq.\ (\ref{eq:univquant}) with other parametric values may be solved by graphical or numerical methods.

Equation (\ref{eq:univquant}) may be generalized to cover antisymmetric quantum reduced actions for other one-dimensional QSHJEs with different symmetric potentials.  Equation (\ref{eq:gw}) for $q=\infty$ may be expressed as

\begin{equation}
\cos\left(\frac{J}{4\hbar}\right)  \big[ A \phi(\infty) + B \vartheta(\infty) \big] = \sin\left(\frac{J}{4\hbar}\right) \big[ C \phi(\infty) + D \vartheta(\infty) \big].
\label{eq:genquant}
\end{equation}

\noindent  The set $\{\phi(q),\vartheta(q)\}$ for $E<V(\pm \infty)$ may be chosen by the superpositional principle so that $\phi$ be antisymmetric and $\vartheta$ be symmetric consistent with the symmetry of $\{\sinh(q),\cosh(q)\}$ where $\phi(\pm \infty)=\pm \infty$ and $\vartheta(\pm \infty) = \infty$ and where $\phi(\pm \infty)/\vartheta(\pm \infty)=\pm 1$ .  If $\cos[J/(4\hbar)]=0$, then $J=4n\pi\hbar=2nh,\ n=1,2,3,\cdots$, the left side of Eq.(\ref{eq:genquant}) is nullified, $\sin[J/(4\hbar)]=\pm1$, and the quantizing condition is

\begin{equation}
D/C = \phi(\infty)/\vartheta(\infty).
\label{eq:genquant1}
\end{equation}

\noindent Likewise, if $[J/(4\hbar)]=0$, then $J=(4n-2)\pi\hbar=(2n-1)h,\ n=1,2,3,\cdots$, the right side of Eq.(\ref{eq:genquant}) is nullified, $\cos[J/(4\hbar)]=\pm1$, and the quantizing condition is

\begin{equation}
A/B=\vartheta(\infty)/\phi(\infty).
\label{eq:genquant2}
\end{equation}

\noindent Bornian probability need not be invoked in this quantizing algorithm.  The prototype quantization by this algorithm is the quantization of the finite square well.  Virtual $\widetilde{J}$s will still produce $\widetilde{W}$s.

A class of microstates that includes the microstate exhibited by Eq.\ (\ref{eq:genquant}) is shown by

\begin{equation}
\cos \left(\frac{J}{4\hbar}\right)  \big[ GA \phi(\infty) + GB \vartheta(\infty) \big] = \sin \left(\frac{J}{4\hbar}\right) \big[ G^{-1}C \phi(\infty) + G^{-1}D \vartheta(\infty) \big]
\label{eq:microstate}
\end{equation}

\noindent where $G$ is a real finite constant.  The quantizing conditions remain $G^{-1}D/(G^{-1}C)=D/C=\phi(\infty)/\vartheta(\infty)$ and $GA/(GB)=A/B=\vartheta(\infty)/\phi(\infty)$.  The normalization, Eq.\ \ref{eq:coefnorm}, is satisfied as $GA \times G^{-1}D - G^{-1}C \times GB = AD-CB =1$.  The microstate exhibited by Eq.\ (\ref{eq:genquant}) has $G=1$.  Further generalizations are beyond the scope of this opus.

As $W$ is monotonic, and as $\widetilde{J}=2(\widetilde{W}|_{q=+\infty}-\widetilde{W}|_{q=-\infty})$ is continuous with $\widetilde{E}$ for a given microstate, it follows that $\widetilde{W}(\{\widetilde{W},\partial_q \widetilde{W},{\partial^2}_q \widetilde{W}\}|_{q=q_0};\widetilde{E},q)$ is a microstate-dependent continuous function of $\widetilde{E}$ for $-\infty < q < +\infty$.

\subsection{Time Parametrization}

Philosophically, as QT is couched in a quantum Hamilton-Jacobi representation, it is set in $\mbox{\boldmath $q$},t$-space, which innately leads to time parametrization and deterministic quantum trajectories, and not in Hilbert space.  Carroll has discussed that standard quantum mechanics in Hilbert space is imprecise by construction and leads to a probabilistic theory for it ignores information that could specify the microstate [\ref{bib:rc2}].  For example, the Heisenberg uncertainty principle is based upon a subset $\{\mbox{position,momentum}\}|_{t=t_0}$ of the set of initial values. $\{W,\partial_q W,{\partial^2}_q
W\}|{q=q_0}$ or $\{q,\dot{q},\ddot{q}\}|_{t=t_0}$, which is necessary and sufficient to specify a unique solution, $W$, to the QSHJE [\ref{bib:fpl13}].

We may now substantiate time parametrization by Jacobi's theorem, Eq.\ (\ref{eq:jacobitheorem}), the Hamilton-Jacobi transformation equation yielding $\tau$.  We apply Jacobi's theorem to the particular $W_1(0,\hbar k,0;E,q)$ for the square well to generate quantum motion in the classically allowed region, $-a<q<a$, to render by Eqs.\ (\ref{eq:jacobitheorem}) and (\ref{eq:gsw1})

\begin{equation}
t-\tau_1 = \frac{\partial W_1(0,\hbar k,0;E,q)}{\partial E} = \frac{\partial \hbar k q}{\partial E} =  \frac{mq}{\hbar k}, \ \ \ |q|<a
\label{eq:jtepsilon}
\end{equation}

\noindent where the epoch $\tau_1=0$.  The relationship for the initial values $\{W_1,\partial_qW_1,{\partial^2}_qW_1\}|_{q=0} = \{0,\hbar k,0\}|_{q=0}$ between $t$ and $q$ for $|q|<a$ is linear and mimics the classical relationship.

In the classically forbidden region, $|q|>a$, time parametrization is generated for quantum motion outside the square well, $q > a$, by applying Jacobi's theorem to $W_2$, Eq.\ (\ref{eq:gw2}), to generate

\begin{eqnarray}
t-\tau_2 = \frac{\partial W_2}{\partial E} & = & \hbar \frac{(\partial_E \phi_2)\vartheta_2 - \phi_2(\partial_E \vartheta_2)}{(A^2_2 + C^2_2) \phi^2_2 + 2(A_2 B_2+C_2 D_2)\phi_2\vartheta_2 + (B^2_2 + D^2_2) \vartheta^2_2}
                                                              \label{eq:swt21}\\
                                         & = & \frac{m(q-a)}{\hbar \kappa \left( \frac{k}{\kappa} \sinh^2[\kappa(q-a)] + \frac{\kappa}{k} \cosh^2[\kappa(q-a)] \right) }
                                                               \label{eq:swt22}
\end{eqnarray}

\noindent where $\tau_2 = ma/(\hbar k)$ by Eq.\ (\ref{eq:jtepsilon}) to account for the transit time inside the square well from $q=0$ to the potential step at $q=a$. In the classically forbidden region, time parametrization is nonlocal.  (Quantum mechanics is nonlocal.)  Equation (\ref{eq:swt21}) presents the generic case for the square well.  Equation (\ref{eq:swt22}) presents the specific case for the initial values specified by $\{W,\partial_q W,{\partial^2}_q W\}|_{q=0} = \{0,\hbar k,0\}$.  Time at the potential step is given by Eq.\ (\ref{eq:swt22}) as $t|_{q=a} = \tau_2$.  Time has the same value in the infinite limit, $\lim_{q \to \infty}t|_q = \tau_2$, also by Eq.\ (\ref{eq:swt22}).  Time in the classically forbidden region increases until it reaches a maximum time, at $\partial t/\partial q  = 0$, whose location is given implicitly by the value of $q$ that solves the transcendental hyperbolic equation

\[
\frac{k}{\kappa} \sinh^2[\kappa(q-a)] + \frac{\kappa}{k} \cosh^2[\kappa(q-a)] - \kappa(q-a) \left(\frac{\kappa^2 + k^2}{k\kappa}\right) \sinh[2\kappa(q-a)] = 0.
\]

\noindent After time reaches its maximum, $t_{\mbox{\scriptsize max}}$, time begins retrograde motion as it monotonically recedes to $\tau_2$ as $q \to \infty$ consistent with Eq.\ (\ref{eq:swt22}), which manifests nonlocality [\ref{bib:fp37a}].  This is consistent with the $\psi$-based Hartman-Fletcher effect for tunneling [\ref{bib:hartman},\ref{bib:fletcher}] where dwell times within thick potential barriers decrease with thickness. The monotonic decrease in time after $t_{\mbox{\scriptsize max}}$ with increasing $q-a$ explains why the Hartman effect is confined to thick barriers. From the inequality implied by the denominator on the right side of Eq.\ (\ref{eq:swt22}),

\[
\left( \frac{k}{\kappa} \sinh^2[\kappa(q-a)] + \frac{\kappa}{k} \cosh^2[\kappa(q-a)] \right) > \frac{\kappa}{k} \cosh^2[\kappa(q-a)] > \frac{\kappa}{k},\ \ \mbox{for}\  q>a,
\]

\noindent it follows that $\dot{q}>\hbar k/m$ in the classically forbidden region and is in agreement with Olkhovsky and Racami [\ref{bib:olkrec}].  The dwell time by Eq.\ (\ref{eq:swt22}) in the classically forbidden region between $a$ and finite $q>a$, that is the value $t(q)-t(a)$ where $a<q<\infty$, decreases with increasing $\kappa$ consistent with tunneling dwell times of Barton [\ref{bib:barton}].  Nevertheless, the conjugate momentum in the classically forbidden region, Eq.\ (\ref{eq:gcm2}), remains finite positive for all finite $q \ge a$.  This ensures that $W_2$ does not become a constant in the finite classically forbidden region, $a<|q|<\infty$, consistent with QEP.  However, QEP considers that trajectories for localized particles with a defined velocity cannot be derived from the QSHJE [\ref{bib:fm3}].

Note that a group velocity for a wave packet has much in common with Jacobi's theorem [\ref{bib:fm3},\ref{bib:fp37a}].  However, Jacobi's theorem does not have the limitations of a group velocity for a wave packet [\ref{bib:fp37a}].  Jacobi's theorem is applicable to widely spread spectra with great divergence in their amplitude.  Jacobi's theorem may render a quantum trajectory through a domain where a wave packet loses its integrity.

\subsection{Cycle Time}

As the quantum trajectory by Eq.\ (\ref{eq:swt22}) transits the classically forbidden region from $q=a$ to $q=\infty$ in nil time, the period of oscillation, $T$, for the set of initial values $\{0,\hbar k,0\}|_{q=0}$ is the same as classical period, $T_{\mbox{\scriptsize classical}}$, for

\begin{equation}
T/4 = \tau_2 + \frac{-m\infty}{\hbar \kappa \left( \frac{k}{\kappa} \sinh^2(\kappa\infty) + \frac{\kappa}{k} \cosh^2(\kappa\infty) \right)} = \tau_2.
\label{eq:swoscillation}
\end{equation}

\noindent The period of oscillation for the finite square well with initial values $\{0,\hbar k,0\}|_{q=0}$ is $T = 4\tau_2 = 4ma/(\hbar k)$. The virtual quantum period of oscillation, $\widetilde{T}=4\widetilde{\tau}_2$  for the finite square well with initial values $\{0,\hbar k,0\}|_{q=0}$ with virtual energies is also equal to the classical period of oscillation by Eqs.\ (\ref{eq:swt21}) and (\ref{eq:swt22}).  For the finite square well, the quantum period of oscillation innately obeys the Bohr correspondence principle for all energies, eigenvalue or virtual.

\section{Findings and Conclusions}

\subsection{Findings}
The phenomenological QSHJE and SSE have {\itshape ab initio} solutions, $W$ and $\psi$ respectively, even if energy is not an eigenvalue. For completeness, the solutions $W$ and $\psi$ mutually imply each other [\ref{bib:fpl9}].

The QT algorithm for quantum trajectories uses the solution for the quantum reduced action as an input for Jacobi's theorem. The QSHJE and Jacobi's theorem are well posed in the Hadamard sense for (1) solutions exist even for non-eigenvalues of energy, (2) their solutions are unique, and (3) their solutions are Lipschitz continuous.  The quantum reduced action, a solution of the QSHJE, is sufficiently Lipschitz continuous to be differentiated by Jacobi's theorem to render, in turn, time parametrization that has Lipschitz continuity.

Energy in the quantum reduce action, the generator for quantum motion, is considered a to be a variable by Eq.\ (\ref{eq:sw}).  After a well posed application, such as Jacobi's theorem, the resultant of the application may be evaluated at a prescribed energy.

QEP and QT have much in common.  Both have a common Hamilton-Jacobi foundation set in $\mbox{\boldmath $q$},t$-space rather than Hilbert space and solve the same QSHJE for the same quantum reduced action $W$ that has a M\"{o}bius transformation character.  For both, the initial values $\{W,\partial_q W,{\partial^2}_q W\}|_{q=q_0}$ are necessary and sufficient to solve the third-order partial differential QSHJE in one dimension for a unique quantum reduced action.  Both agree that the quantum reduced action contains more information than $\psi$ [\ref{bib:prd25}--\ref{bib:fm},\ref{bib:rc2}].  This additional information for the particular solution $W(E,\{W,\partial_q W,{\partial^2}_q W\}|_{q=q_0};q)$ is common to both QEP and QT.  Both QEP and QT agree that this additional information renders microstates (M\"{o}bius states) for bound state $\psi$.  Both agree that Jacobi's theorem does not generally render localized quantum trajectories [\ref{bib:fm3}], albeit QT does render nonlocal trajectories.  Both also agree that the conjugate momentum is generally not the mechanical momentum, $\partial_q W \ne m\dot{q}$, for the Schwarzian derivative is generally not a conservative potential  [\ref{bib:prd26},\ref{bib:fm}].

Under QT, formulas for quantizing energy may be developed without using the fact that the bound-state $\psi$ is $L^2(\mathbb{R})$.

For the linear harmonic oscillator, the relationship between $J\omega = 2\pi (E + 0.5 \hbar \omega)$ is valid for eigenvalues $J$ and $E$ regardless of the initial values $\{W,\partial_q W,{\partial^2}_q W\}|_{q=q_0}$.  The relationship between $\widetilde{J}$ and $\widetilde{E}$ is dependent upon the particular initial values $\{\widetilde{W},\partial_q \widetilde{W},{\partial^2}_q \widetilde{W}\}|_{q=q_0}$, cf.\ Tables 1 and 2.

The Bohr correspondence principle applies even to virtual states.  As $\widetilde{E}$ increases by increments of $\hbar \omega$ for the linear harmonic oscillator, the numerical behavior of $\widetilde{J}$ converges to the classical behavior.  This is consistent with the Bohr correspondence principle as exhibited by Table 3 even for a $\widetilde{J}$ with an $\widetilde{E}$ midway between its neighboring eigenvalues $E$s. Transit times for a quarter-cycle for the linear harmonic oscillator for virtual states with $\widetilde{E}s$ midway between their neighboring eigenvalues $E$s  converge to the classical value with increasing $\widetilde{E}$ consistent with the Bohr correspondence principle as shown by Table 4.  Transit time for a quarter-cycle for the finite square well for any virtual state energy is the classical time as shown by Eq.\ (\ref{eq:swoscillation}).

Non-eigenvalue energies have been shown in \S2 and \S3 to yield well-behaved $\widetilde{J} \equiv 2 \widetilde{W}|^{+\infty}_{q=-\infty}$.  For a given set of initial values, $\{W,\partial_q W,{\partial^2}_q W\}_{q=q_0}$, of the QSHJE, these virtual quantum reduced actions are sufficiently well-behaved with regard to energy to find quantized action variables by shooting techniques [\ref{bib:st}] as initially shown in unpublished Ref.\ \ref{bib:prvda25} and substantiated herein with greater precision, cf. Tables 1 and 2.

QT can derive quantization formulas of the wave representation for the finite square well as demonstrated in alternate forms by Eqs.\ (\ref{eq:qsymm}), (\ref{eq:qanti}), (\ref{eq:gwquant}), and (\ref{eq:univquant}).  This may be generalized for other symmetric potentials, Eqs.\ (\ref{eq:genquant1}) and (\ref{eq:genquant2}). These quantizing equations generally are intrinsic, transcendental equations whose solutions tacitly assume continuity of $\widetilde{W}$ with $\widetilde{E}$ consistent with Jacobi's theorem.   QT can derive the number of bound states for the finite square well, cf. Eq.\ (\ref{eq:maxni}).  By implication, QEP should do so too.

\subsection{Conclusions}

For an open universe, Jacobi's theorem, being well posed, dose render time parametrization that has Lipschitz continuity.

As the QSHJE problem is well posed, and as $W$ must have monotonic behavior [{\ref{bib:fm}], the numerical analyses problem is simplified.  For closed-form analyses, choosing a bound-state solution set $\{\psi,\vartheta\}$ of the SSE so that
$\lim_{q \to \pm \infty} \psi/\vartheta = \pm 1$ simplifies finding the quantization formulas for symmetric potentials, Eqs. (\ref{eq:genquant})--(\ref{eq:genquant2}).

While QEP and QT find the same energy quantization for bound states, they have their own criterion for finding energy quantization.  QEP's criterion for energy quantization is that $\psi_{\mbox{\scriptsize bound}}$ be $L^2(\mathbb{R})$ compliant.  And the QT criterion for energy quantization is implied by Milne quantization, $J_{\mbox{\scriptsize bound}} = 2n \pi \hbar,\ n=1,2,3,\cdots$.  Both approaches for quantization are consistent with each other.   While QEP does not assume any Copenhagen axioms, it finds that conservation of Bornian probability is still consistent with the $L^2(\mathbb{R})$ character of bound states [\ref{bib:fm3}].  In contrast, QT renders deterministic quantum trajectories that are incompatible with ascribing a probability interpretation to $\psi$. (Lest we forget, E.\ Schr\"odinger opposed a Bornian probability amplitude for his $\psi$.)

Faraggi has interpreted that under QEP the Schwarzian derivative in the QSHJE, Eq.\ (\ref{eq:qshje}) would manifest an internal structure for elementary particles  [\ref{bib:aef13}].  Under QT, the finite $\hbar^2$ factor in the Schwarzian derivative term would preclude quantum mechanics from using an infinitesimal test charge to establish the field associated with a potential, $V$ [\ref{bib:prd25}].  An infinitesimal test charge does not distort the field [\ref{bib:jdjackson}].  These two interpretations of the Schwarzian derivative term are compatible with each other.

The differences between QEP and QT that were developed herein may be teleological regarding spatial compactification.  QEP supposes that spatial compactification would suggest that time is probabilistic [\ref{bib:fm3}] and that the propagation of gamma rays over astrophysical distances may be affected [\ref{bib:fm5}].  QT supposes an open universe which would allow the use of information (initial values) not available in the $\psi$-representation of quantum mechanics to imply time parametrization from deterministic, nonlocal quantum trajectories.

QEP forbids local trajectories except as a semi-classical approximation [\ref{bib:fm3}]. QT allows nonlocal quantum trajectories.  A quantum trajectory may have segments of temporal retrograde motion interspersed between segments of temporal forward motion.   Under the Stueckelberg hypothesis [\ref{bib:stueckelberg}], segments of retrograde motion for a particle manifest temporal forward motion of its antiparticle. The particle-antiparticle pairs are non-endoergically created and non-exoergically annihilated respectively  at temporal minima and maxima of the nonlocal quantum trajectory in $q,t$-space [\ref{bib:fp37a}].

\bigskip

\small

\centerline{\bfseries Acknowledgement}

\medskip

I heartily thank A.\ E.\ Faraggi and M.\ Matone for their invited, helpful comments on an earlier version, especially on M\"{o}bius transformations.  Although we differ on compactification, their correspondence has always been cordial.  I also thank one of the referees whose incisive comments contributed to significant improvements in the paper.

\bigskip

\noindent {\bfseries REFERENCES}

\begin{enumerate}\itemsep -.06in\item

\label{bib:prd25} Floyd, E.\ R.: Born-Sommerfeld Quantization with the effective action variable. Phys.\ Rev.\ D {\bfseries 25}, 1547 (1982).

\item \label{bib:prvda25} Floyd, E.\ R.: Physics Auxiliary Publication Service to Ref.\ \ref{bib:prd25}, PAPS PRVDA 25-1547-20; order free pdf copy from
$<$aip.org/epaps$>$ citing PAPS Number and journal reference (Reference \ref{bib:prd25} above).

\item \label{bib:prd26}  Floyd, E.\ R.: Modified potential and Bohm's quantum potential.   Phys.\ Rev.\ D {\bfseries 26}, 1339 (1982).

\item \label{bib:prd29} Floyd, E.\ R.: Arbitrary initial conditions for hidden variables.  Phys.\ Rev.\ D {\bfseries 29,} 1842 (1984)

\item \label{bib:prd34} Floyd, E.\ R.: Closed form solutions for the modified potential.  Phys.\ Rev.\ D {\bfseries 34}, 3246 (1986).

\item \label{bib:fpl9} Floyd, E.\ R.:   Where and why the generalized Hamilton-Jacobi representation describes microstates of the Sch\"{o}dinger wave function.  Found.\ Phys.\ Lett.\ {\bf 9}, 489 (1996), quant-ph/9707051.

\item \label{bib:rc} Carroll, R.: Some remarks on time, uncertainty, and spin. J.\ Can.\ Phys. {\bfseries 77}, 319 (1999), quant-ph/9903081.

\item \label{bib:fpl13} Floyd, E.\ R.:  Reflection time and the Goos-H\"{a}nchen effect for reflections from a semi-infinite rectangular barrier. Found.\ Phys.\ Lett.\ {\bfseries 13}, 235 (2000), quant-ph/9708070.

\item \label{bib:fm}  Faraggi, A.\ E., Matone, M.: The equivalence postulate of quantum mechanics. Int.\ J.\ Mod.\ Phys.\ A {\bfseries 15}, 1869 (2000), hep-th/9809127.

\item \label{bib:ijmpa15} Floyd, E.\ R.: Classical limit of the trajectory representation of quantum mechanics, loss of information and residual indeterminacy.    Int.\ J.\ Mod.\ Phys.\ A {\bfseries 15}. 1363 (2000), quant-ph/9907092.

\item \label{bib:wyatt} Wyatt, R.\ E.: Quantum Dynamics with Trajectories: Introduction to Quantum Hydrodynamics, (Springer, New York, 2005) pp 357--363.

\item \label{bib:pr35} Milne, W.\ E,: The numerical determination of characteristic numbers. Phys.\ Rev.\ {\bfseries 35}, 863 (1930).
(Cambridge, New York, 2007) pp 959--64

\item \label{bib:pl450} Faraggi, A.\ E., Matone, M.: Quantum mechanics from an equivalence principle.  Phys.\ Lett.\ B {\bfseries 450}, 34 (1999), hep-th/9705108.

\item \label{bib:fm3} Faraggi, A.\ E., Matone, M.: Energy Quantisation and Time Parameterisation.  Eur.\ Phys.\ J.\ C {\bfseries 74}, 2694 (2014), arXiv:1211.0798v2.

\item \label{bib:aef13} Faraggi, A.\ E.: The quantum closet.  In: Dobrev (ed): Lie Theory and its Applications in Physics, Springer Proceedings in Mathematics \& Statistics, Vol.\ 111, pp 541--9 (Springer, Berlin Heidelburg New York, 2014).   arXiv:1305.0044.

\item \label{bib:fm4} Faraggi, A.\ E., Matone, M.: The M\"{o}bius symmetry of quantum mechanics. arXiv:1502.04456.

\item \label{bib:fm5} Faraggi, A.\ E., Matone, M.: Hamilton-Jacobi meet M\"{o}bius. arXiv:1503.01286.

\item \label{bib:jh} Hadamard, J.: Sur les probl\`emes aux d\'eriv\'ees partielles et leur signification physique.  Bull.\ Univ.\ Princeton {\bfseries 13} 49--52 (1902).

\item \label{bib:ik} Isaacson, E.\, Keller, H.\ B.: Analysis of Numerical Methods, (John Wiley, New York, 1966) pp 1, 22, 23, 27, 139, 444.

\item \label{bib:st} Press, W.\ H., Teukolsky, S.\ A.,  Vetterling, W.\ T., Flannery, B.\ P.: Numerical Recipes: The Art of Scientific Computing, 3rd ed.

\item \label{bib:tl}  Tipler, P., Llewellyn, R.: Modern Physics, 5th ed.\ (W.\ H.\ Freeman and Co., New York, 2008) pp 160--1.

\item \label{bib:eh} Hille, E.: Ordinary Differential Equations in the Complex Plain. (Dover: Mineola, NY, 1976) pp 374--401.

\item \label{bib:arf} Forsyth, A.\ R.: A Treatise on Differential Equations, 6th ed. (Macmillan, London, 1929) pp. 104--5, 320--2.

\item \label{bib:hm}  Hecht, C.\ E., Mayer, J.\ E.: Extension of the WKB equation. {\itshape Phys.\ Rev.} {\bfseries 106} 1156--60 (1953).

\item \label{bib:vigierconf} Floyd, E.\ R.: The philosophy of the trajectory representation of quantum mechanics. In: Amoroso, R.\ L., Hunter, G.,
Kafatos, M., Vigier, J.-P. (eds.) {\itshape Gravitation and Cosmology: From the Hubble Radius to the Planck Scale; Proceedings of a Symposium in Honour of the 80th Birthday of Jean-Pierre Vigier}, (Kluwer Academic, Dordrecht, 2002) pp 401-408, extended version promulgated as quant-ph/00009070.

\item \label{bib:fpc} Souradeep, T.,  Pogsyan, D., Bond, J.\ R.:  Probing cosmic topology using CMB anistropy. In: V$\hat{\mbox{a}}$n, J.\ T.\ T., Graud-H$\acute{\mbox{e}}$raud, Y., Bouchet F., Damour T., Mellier Y.\ (eds.) Fundamental Parameters in Cosmology, (Editions Fronti$\grave{\mbox{e}}$res, Paris, 1998) pp 131--3.

\item \label{bib:bo} Bender, C.\ M., Orszan, S.\ A.: Advanced Mathematical Methods for Scientists and Engineers. (McGraw-Hill, New York, 1978), pp. 171--8.

\item \label{bib:dwight} Dwight, H.\ W.:  Table of Integrals and Other Mathematical Data, 4th ed. (Macmillan, New York, 1961) \P401.2.

\item \label{bib:eisberg}  e.g., Eisberg, R.\ M.: Fundamentals of Modern Physics. (John Wiley, New York, 1961) pp. 239--51.

\item \label{bib:rc2} Carroll, R.: Quantum Theory, Deformation and Integrability, (Elsevier, Amsterdam, 2000) pp. 50--6.

\item \label{bib:fp37a}  Floyd, E.\ R.: Interference, reduced action and trajectories. Found.\ Phys.\ {\bfseries 37}, 1386 (2000), quant-ph/0605120v3.

\item \label{bib:hartman} Hartman, T.\ E.: Tunneling of a wave packet. J.\ Appl. Phys. {\bfseries 33}, 3427 (1962).

\item \label{bib:fletcher} Fletcher, J.\ R.: Time delay in tunnelling through a potential barrier. J.\ Phys. {\bfseries C 18}, L55 (1985)

\item \label{bib:olkrec} Olkhovsky, V.\ S., Racami, E.:  Recent developments in the time analysis of tunneling processes. Phys.\ Rep. {\bfseries 214}, 339 (1992).

\item \label{bib:barton} Barton, G.: Quantum mechanics of the inverted oscillator potential. Ann.\ Phys.\ (NY) {\bfseries 166}, 339 (1986).

\item \label{bib:jdjackson} Jackson, J.\ D.: Classical Electrodynamics. (Wiley, New York, 1962) p. 2.

\item \label{bib:stueckelberg} Stueckelberg, E.\ C.\ G.: La signification du temps propre en mécanique ondulatoire. Helv.\ Phys.\ Acta. {\bfseries 14} 51 (1941).

\end{enumerate}

\end{document}